\documentclass[11pt,a4paper]{article}
\usepackage[utf8]{inputenc}
\usepackage{amsmath}
\usepackage{placeins}
\usepackage{url}
\usepackage{natbib}
\usepackage{color,xcolor,setspace,geometry}
\setcitestyle{authoryear,open={(},close={)}}
\usepackage{cite}
\usepackage{amsfonts}
\usepackage[normalem]{ulem}
\usepackage{pifont}

\usepackage{arydshln}

\usepackage{tikz}
\usetikzlibrary{shapes}
\usetikzlibrary{plotmarks}
\usetikzlibrary{tikzmark,matrix,calc}
\usetikzlibrary{arrows.meta,calc,decorations.markings,math,arrows.meta}

\usepackage{amssymb}
\usepackage{multirow}
\usepackage{graphicx}
\usepackage{soul}
\soulregister\citep7
\soulregister\cite7
\soulregister\citet7
\soulregister\citealp7
\soulregister\ref7
\usepackage{rotating}
\usepackage{setspace}
\usepackage{threeparttable}
\usepackage{authblk}
\usepackage{subfigure}

\newcommand{\ts}{\textsuperscript}
\usepackage{appendix}

\begin{document}
\begin{spacing}{1.5}	

\title{Dynamic Stochastic Inventory Management\\ in E-Grocery Retailing}

\author[$\star$ $\ddag$]{David Winkelmann}
\author[$\star$]{Matthias Ulrich}
\author[$\star$]{Michael Römer}
\author[$\star$]{Roland Langrock}
\author[$\star$]{Hermann Jahnke}

\affil[$\star$]{Bielefeld University, Universit\"atsstrasse 25, Bielefeld, Germany.}
\affil[$\ddag$]{Corresponding author: david.winkelmann@uni-bielefeld.de}
\maketitle
\noindent

\begin{abstract}
E-grocery retailing enables ordering products online to be delivered at a future time slot chosen by the customer. This emerging field of business provides retailers with very large and comprehensive new data sets, yet creates several challenges for the inventory management process. For example, the risk of a single item's stock-out leading to a complete cancellation of the shopping process is higher in e-grocery than in traditional store retailing. As a consequence, retailers aim at very high service level targets to provide satisfactory customer service and to ensure long-term business growth. When determining replenishment order quantities, it is therefore of crucial importance to precisely account for the full uncertainty in the inventory process. This requires predictive and prescriptive analytics to (1) estimate suitable underlying probability distributions to represent the uncertainty caused by non-stationary customer demand, shelf lives, and supply, and to (2) integrate those forecasts into a comprehensive multi-period optimisation framework. In this paper, we model this stochastic dynamic problem by a sequential decision process that allows us to avoid simplifying assumptions commonly made in the literature, such as the focus on a single demand period. As the resulting problem will typically be analytically intractable, we propose a stochastic lookahead policy incorporating Monte Carlo techniques to fully propagate the associated uncertainties in order to derive replenishment order quantities. This policy naturally integrates probabilistic forecasts and allows us to explicitly derive the value of accounting for probabilistic information compared to myopic or deterministic approaches in a simulation-based setting. In addition, we evaluate our policy in a case study based on real-world data where the underlying probability distributions need to be estimated from historical data and explanatory variables. Our findings illustrate the importance of considering the dynamic stochastic environment for a cost-effective inventory management process.

\end{abstract}

\section{Introduction}
\label{sec1}

Electronic groceries (e-groceries) comprise click-and-collect services and attended home delivery, i.e.\ the additional service of delivering from the retailer to customers \citep{saunders2018}. The associated increase in complexity and operational costs of logistic processes compared to brick-and-mortar retailing motivates the research question of this paper. From the supply side, e-grocery retailers face low profit margins and high fulfilment costs \citep{akkerman2010quality, hubner2013demand}. These fulfilment costs typically cannot be fully passed on to price-sensitive customers via delivery fees. Although this constitutes a challenge, the accelerated growth of e-grocery demand as a result of the coronavirus (COVID-19) pandemic motivates further investments in already scaled-up capacities (\citealp{mckinsey2021}; for the dynamic growth and the importance especially of click-and-collect e-grocery see, e.g., \citealp{siawsolit2021offsetting}). Improving inventory processes in e-grocery retailing may help to reduce fulfilment costs. However, given the stochastic environment it operates in, this requires decision support when determining replenishment order quantities. Management strategy aims at achieving demand satisfaction at a high target level. Consequently, the full uncertainty in the inventory process must be recognised and suitable estimation methods for an extremely right hand quantile of the demand distribution are required. Therefore, this paper addresses two closely connected aspects of operative decision making in e-grocery retailing stemming from the business environment under consideration: (1) formulating an adequate inventory management model, and (2) integrating all available information on the stochastic variables impacting inventory levels into the decision-making process.

In practice, most retailers offer stock keeping units (SKUs) with a shelf life of multiple demand periods. As a result, excess inventory can be sold in the following demand period(s) and thus affects the replenishment order decisions in those periods \citep{kim2015optimal}. In addition to these dynamic inter-period dependencies, retailers are faced with a convolution of distributions for multiple stochastic variables impacting the inventory level such as demand, shelf lives, and the quantity delivered by the supplier. In our case, the underlying probability distributions are non-stationary and depend on exogenous variables (features) such as weather conditions as well as endogenous decisions such as the price for the SKU, which affects its demand. These uncertainties are typically amplified by a lead time of multiple days. Hence, the costs resulting from a given order decision are uncertain. For supporting operative order decision making, fully capturing the business environment within an adequate inventory model thus results in a dynamic stochastic inventory problem --- which is notoriously difficult to solve (see \citealp{bijvank2011lost}).

This task becomes even more challenging as the distributions of the various uncertain variables impacting the stochastic inventory process are typically unknown to the decision maker. Retailers therefore have to deal with two key challenges that make it difficult to integrate available information when determining optimal replenishment orders (\citealp{Fildes2019}). First, the underlying probability distributions or their parameters need to be estimated from historical data, and perhaps features, using suitable prediction methods. Second, retailers need to adequately incorporate those forecasts and other results on the various sources of uncertainty from this first phase of analysis into the decision-making process (\citealp{Raafat1991}, \citealp{Silver1998}). Due to the online nature of e-grocery retailing, there is comprehensive data pertaining to customer behaviour available. In our case, this opens the path to using approaches from \textit{descriptive} and \textit{predictive analytics} to solve the first task. The utilisation of insights generated by these two steps in the decision making process is referred to as \textit{prescriptive analytics} (see \citealp{lepenioti2020prescriptive}, for a literature review).

The existing literature emphasises the value of new types of data available in e-grocery compared to brick-and-mortar retailing, such as information on unbiased customer preferences given by uncensored demand data \citep{Ulrich2018} or orders by customers for future demand periods known in advance \citep{siawsolit2021offsetting}. Indeed, these data enhance the quality of demand distribution estimations that need to be taken into account when determining replenishment order quantities. However, recent data-driven approaches for inventory management mostly focus on modelling uncertainty in customer demand only, making restrictive assumptions concerning other sources of uncertainty. For example, \citet{Xu2021} consider shelf lives to be fixed at a single demand period as within the newsvendor model. In this paper, we address these limitations by proposing a flexible multi-period inventory management framework that explicitly enables to consider perishable goods with a stochastic shelf life of multiple periods. While approaches based on a multi-period newsvendor setting assume a fixed shelf life (see e.g.\ \citealp{kim2015optimal}) or backordering (see e.g.\ \citealp{zhang2020optimal}), our lost sales model can represent shelf-life either via a fixed value or using a probability distribution, and additionally allows to incorporate the risk of potential delivery shortfalls. Thus, we are able to take into account all relevant stochastic variables, namely demand, shelf lives of SKUs, and delivery shortfalls using suitable probability distributions to allow for a data-driven inventory management process meeting the requirements of a real-world e-grocery retailing business case. 

Typically, there is no closed-form solution of the optimal policy for inventory control problems, particularly in case of lost sales and a lead time of multiple days \citep{boute2022deep}. In our problem, we also fully accommodate the various uncertainties described above. This renders it even more challenging to derive optimal replenishment order quantities in this dependent multi-period setting. Therefore, we propose a Monte Carlo-based approximate dynamic programming approach that determines the replenishment order decisions minimising the expected costs for a set of sample trajectories spanning a given lookahead horizon. An advantage of this approach, which, following the terminology proposed by \citet{Powell2019}, can be characterised as a \textit{stochastic lookahead policy}, is that it allows to integrate the full distributional information of all stochastic variables available to the decision-makers.
The potential of such approximate numerical methods for complex inventory control problems is shown by \citet{boute2022deep}. From a modelling perspective, a key benefit of these approaches is that they do not require stationarity assumptions but naturally integrate time-dependent probabilistic forecasts such as those suggested by \citet{Ulrich2018} for customer demand in e-grocery retailing. A similar approach has previously been applied to routing problems in a stochastic dynamic environment \citep{Brinkmann2019, soeffker2022stochastic}.

For the evaluation of the lookahead policy proposed in this paper, we first test the policy in a simulation-based setting, where we can consider the benefit of incorporating full uncertainty information in isolation, i.e.\ without the additional noise induced by the need to estimate the relevant probability distributions. After deriving replenishment order quantities based on the newsvendor model and a deterministic approach as a benchmark, we apply the stochastic lookahead policy to the same data set. This allows us to assess the benefit of (1) using this approach in the first place, instead of the myopic newsvendor model, and (2) using probability distributions instead of deterministic expected values for the stochastic variables affecting the replenishment order decision process. In these simulations, we further discuss the sensitivity of the results with respect to the specification of different model parameters. Second, we evaluate our policy in a case study using real-life data from a European e-grocery retailer, with the additional challenge that the stochastic variables' probability distributions need to be estimated from historical data and vary over time. The data is used to generate probabilistic forecasts which are fed into the stochastic lookahead policy. The practical applicability of our approach is further demonstrated by comparing it to a parametric decision rule used in practice by the e-grocery retailer considered.

\section{E-grocery retailing and related literature}
In this paper, we consider the business case of a European e-grocery retailer who aims at optimising replenishment order quantities while facing several sources of uncertainty in the inventory management process. This section provides an overview on related literature dealing with inventory management problems as well as an introduction into the specifics of e-grocery retailing business. We start by giving a review on basic inventory models, followed by approaches attempting to take into account challenges for inventory modelling in retailing practice, such as stochastic supply and shelf lives. Finally, we introduce specifics of e-grocery retailing.

\subsection{Basic approaches to modelling inventory management problems}
The consideration of inventory management problems has a long tradition in the literature (see e.g.\ \citealp{silver1981operations}). In general, inventory management deals with the questions on how often, when and which quantity to replenish. Real-world problems differ a lot in the underlying circumstances, such as the type of products, the uncertainty in the inventory management process as well as the cost structure. In the past, the literature mostly focused on simple decision policies for determining replenishment order quantities (\citealp{Heyman2004}).
The two basic models, both associated with very limiting assumptions, are the EOQ model and the newsvendor model. Back in 1913, the EOQ model was the first to provide decision support for companies when it comes to the question of replenishment order quantities \citep{erlenkotter1990ford} and still forms the basis for more recent approaches (see e.g.\ \citealp{alinovi2012reverse}). However, it is restricted by a constant and known demand rate as well as an infinite shelf life of SKUs. On the other hand, the newsvendor model states the classical inventory management model to determine the cost-optimal inventory level in case of stochastic customer demand (\citealp{Silver1998}, \citealp{Zipkin2000}). Again, restrictive assumptions have to be accepted: In its basic version, the model considers independent demand periods, i.e.\ a shelf life of one demand period only.

In case of stochastic customer demand, two situations can arise at the end of a demand period: (1) demand exceeds the inventory level leading to either lost sales or backordering of orders and (2) excess inventory. While most inventory management models consider the case of backordering, especially in the practice of grocery retailing the assumption of lost sales is more realistic, leading to models that are in general more difficult to solve (see \citealp{bijvank2011lost} for a review on inventory models with lost sales). Under the assumptions of the newsvendor model, the cost-optimal inventory level can be derived by the ratio between costs for excess inventory and costs for excess demand. In their review, \citet{qin2011newsvendor} give suggestions for future research based on the newsvendor model, such as the integration of stochastic supply and demand in the same model as well as the introduction of (stochastic) lead times and multi-period models.

More recently, retailers are able to collect comprehensive data at low costs while at the same time, the available computational power has increased. These developments made it possible to design new data-driven approaches for inventory management (see e.g.\ \citealp{Elmachtoub2021, Lee2021, Xu2021}; \citealp{Ralfs2022}). In particular, e-grocery retailing offers opportunities for an accurate analysis of decision policies, as external effects are reduced and the data is typically more informative than in brick-and-mortar retailing, e.g.\ due to the availability of uncensored demand data \citep{Ulrich2018}. However, these approaches incorporating extensive data are again mainly based on the setting of the newsvendor problem with its restrictive assumptions. In their recent review, \citet{boute2022deep} highlight the potential of approximate numerical methods like deep reinforcement learning for complex sequential decision problems such as inventory control  problems with lead times and lost sales. Agreeing with the assessment of these authors, we propose an approximate dynamic programming approach for the e-grocery inventory management problem considered in this paper.

\subsection{Challenges for inventory modelling from retailing practice}
To account for the characteristics of practical problems, several extensions of basic inventory models have been proposed. One crucial matter is the choice of an appropriate probability distribution used for representing random demand as observed by the retailer. Parts of the previous literature rely on simple modelling such as considering a Poisson process for the arrival of customer demand (see e.g.\ \citealp{siawsolit2021offsetting}). While such restrictive assumptions make deriving optimal replenishment order policies easy \citep{bijvank2011lost}, they are not descriptive in many applications. \citet{Ulrich2021}, e.g., based on their real-world e-grocery retailing data, demonstrate the importance of a case-specific estimation of the demand distribution. Especially, the combination of high service level requirements and more complex demand patterns commonly observed in e-grocery retailing favour the use of probability distribution that allow for e.g.\ overdispersion or skewness.

In addition to uncertain customer demand, most SKUs in grocery retailing have a finite shelf life of multiple periods, evoking costs for inventory holding at the end of each period and spoilage costs as well as stock reductions if they are not sold within their shelf life (see e.g.\ \citealp{siawsolit2021offsetting}). The resulting reductions in the inventory level need to be considered in the replenishment order decision. \citet{kim2015optimal} consider the case of a multi-period newsvendor model to allow the consideration of perishable SKUs with a finite shelf life under non-stationary demand, thus extending previous work. Other contributions to the literature discuss the cases of fixed and stochastic shelf lives; see the surveys of \citet{Nahmias1982} and \citet{Raafat1991}. Most literature considering finite shelf lives assumes that the number of sales periods is known (e.g.\ \citealp{Myers1997,Chowdhury2001,Viswanathan2002}) or that each SKUs decays at a constant rate \citet{Kaya2017}. However, for persishable SKUs such as fruits and vegetables, the number of sales periods is more realistically represented by a random variable. The associated probability distribution can be estimated by modelling the decay of the SKUs in the course of time. The rate of decay can be described by a constant fraction of the given inventory or by following a rate that changes according to an underlying function \citep{Raafat1991}, as for example an exponential distribution \citep{Nahmias1982}.

In general, retailers additionally face the risk of supply shortages, e.g.\ due to supply constraints in the distribution channels. This problem is referred to as \textit{random yield} in the literature. Existing supply-uncertainty literature assumes that retailers know their suppliers’ true supply distributions (see e.g.\ \citealp{Yano1995,Grasman2007,Tomlin2009}). \citet{Noori1986} were among the first to address problems where supply and demand are both random, deriving the optimal order quantity for the unconstrained newsvendor problem with random yield. \citet{Parlar1995} allow for non-stationary supply by assuming that supply follows a Bernoulli process, i.e.\ the realisation of no or complete supply.

\subsection{Specifics of e-grocery retailing}
\label{sec:e-grocery}
In e-grocery, customers order groceries online and the retailer directly delivers the purchase from local distribution warehouses to the customer. It differs from traditional store retailing with respect to several characteristics, leading to specific challenges for the optimisation of logistic processes. E-grocery retailers can be broadly categorised into two groups: (1) those who have their roots in classical brick-and-mortar retailing (see \citealp{wollenburg2018bricks} for a review on related logistic processes) and (2) so-called \textit{pure online} grocers \citep{hubner2016retail}. The main challenges e-grocery retailers have to deal with arise from specific questions on warehousing and order picking (see e.g.\ \citealp{winkelmann2022integrated}) as well as the management of delivery slots (see e.g.\ \citealp{wassmuth2023demand}). At the same time, there are specifics regarding the inventory management process. An extensive literature review on logistic strategies in e-grocery retailing is given by \citet{rodriguez2022grocery}.

Opportunities for inventory management optimisation result from new types of data in e-grocery that are not available in traditional store retailing. During the ordering process no in-stock information is available to the customer, which allows to monitor customer preferences and, therefore, yields uncensored demand data. This uncensored demand data is particularly relevant as it allows the retailer to explicitly determine the amount of lost sales, something not possible for check-out data resulting from traditional brick-and-mortar retailing. This enables a more accurate calculation of costs incurred by a specific replenishment order quantity. A further advantage results from the possibility of customers selecting a delivery slot up to fourteen days in advance. This provides information on \textit{known demand}, which equals the customer order quantity for a future delivery period at the time of determining the replenishment order quantity by the retailer. This information can be incorporated into the forecast of demand. Previous literature also considers the case of \textit{advanced demand information} by assuming a higher willingness to pay for shorter lead times \citep{gallego2001integrating} and potential benefits rewarded to customers who are willing to place their orders in advance \citep{siawsolit2021offsetting}.

\begin{figure}[ht]
\centering
\includegraphics[width=0.4\textwidth]{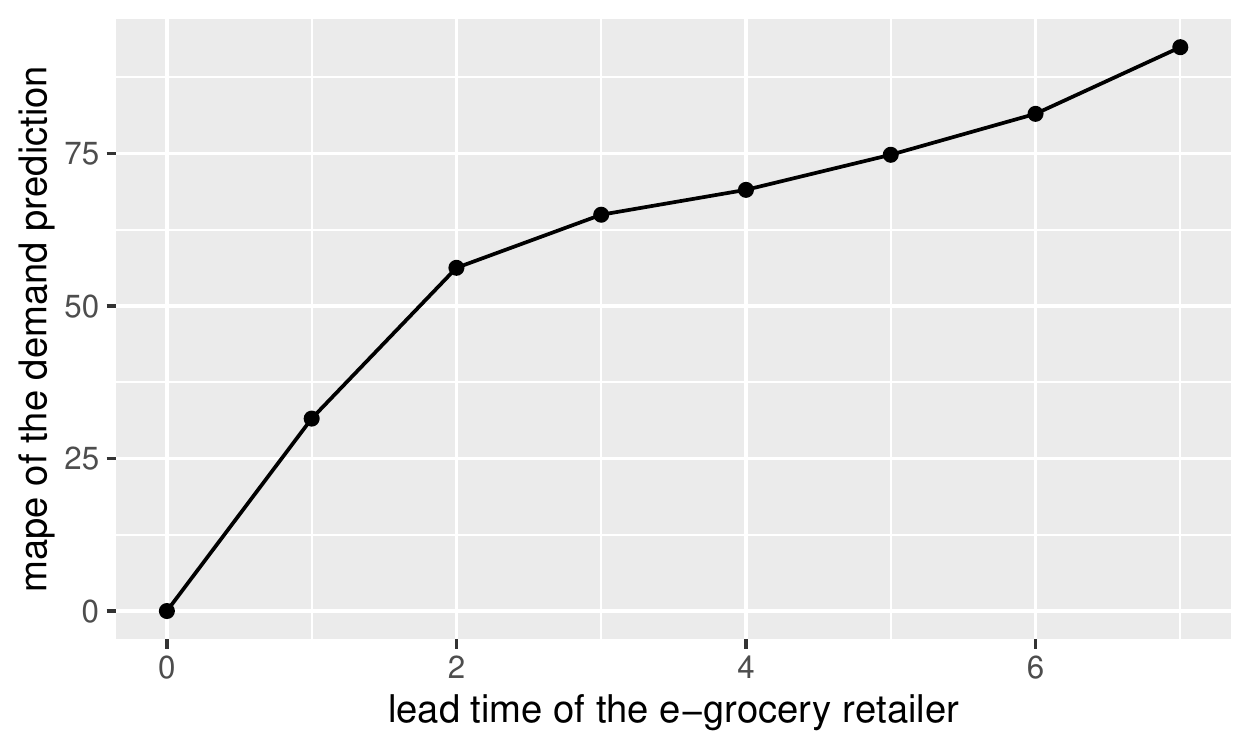}
\caption{Mean average percentage error (mape) as a function of the delivery time of the e-grocery retailer for all SKUs within the categories fruits and vegetables in the demand period January 2019 to December 2019.} \label{figure_leadTime}
\end{figure}

On the other hand, while there are opportunities resulting from the control the retailer exerts over the fulfilment process, picking and delivery increase the time between the instance a replenishment order for an SKU is placed and the final availability to the customer. This longer delivery time reduces the forecasting accuracy of crucial variables, such as the demand distribution for the period under consideration. In particular, features used for the forecast on this distribution, such as the known demand are less informative more days in advance. Using data from the e-grocery retailer under consideration in this paper, Figure~\ref{figure_leadTime} displays the mean average percentage forecast error as a function of the lead time when applying a linear regression for all SKUs within the categories fruits and vegetables in the demand period January 2019 to December 2019. We observe that the mean average percentage error strongly increases for longer lead times, thus implying a decrease in the forecast precision.

At the same time, due to the inconvenience of having to stay at home when the purchase is delivered to the customer, there is a high risk that a customer cancels the whole (virtual) shopping basket during the online purchasing process in case of stock-outs. Note that typically backordering of orders is not possible but unfulfilled demand has to be considered as lost sales. To account for strategic long-term objectives, e-grocery retailers operate with very high service-level targets of 97\% to 99\%, which require to apply complex forecasting methods to appropriately determine an extremely right quantile of the demand distribution \citep{Ulrich2018}. It is therefore crucial for effectively supporting decision making in inventory management to be able to integrate those forecasts when determining replenishment order quantities.

To adapt to the requirements of inventory management in an e-grocery retailing business case introduced above, in this paper, we aim at extending previous literature by considering the closely connected tasks of estimating underlying probability distributions to represent the full uncertainty in a multi-period inventory management problem and integrating those forecasts into an optimisation framework. Our data-driven approach allows for a flexible representation of all relevant underlying uncertainties, e.g.\ by a case-specific selection of the probability distribution for customer demand. To the best of our knowledge, we are the first considering the integration of those three aspects.

\section{Modelling framework and solution approach}
\label{sec:model}

In this section, we introduce a framework for supporting an e-grocery retailer's operative inventory management decision making and propose our solution approach. First, we describe some specifics of the problem for the retailer under consideration. This forms the backdrop for developing an inventory model covering the inventory level dynamics and for deriving a cost function. Here, we explicitly outline how we model the stochastic variables, namely demand, spoilage, and supply shortages, and integrate them into our optimisation framework. Afterwards, we formulate the problem as a sequential decision process and introduce a stochastic lookahead policy that is capable of exploiting the uncertainties via probability distributions when determining replenishment order quantities.

\subsection{Problem of the e-grocery retailer under consideration}
\label{sec:retailer-problem}
This paper attempts to support inventory management decisions of a real-world e-grocery retailer. While the more general characterisation of e-grocery retailing in Section~\ref{sec:e-grocery} also holds for the company under consideration, we will introduce its specific requirements and characteristics in this section.

The assortment of the retailer covers several thousand SKUs from different areas such as fruits, vegetables and meat. The outward distribution process covers two phases. First, at each day there is a supply from national distribution warehouses to local fulfilment centres. In the following, we refer to these warehouses as the supplier. If a customer places a purchase, then this order is served by the dedicated fulfilment centre. This requires the retailer to adequately manage, on a daily basis, the inventory process for a variety of SKUs and several fulfilment centres. Our modelling approach takes the perspective of a single fulfilment centre, where we assume an unlimited storage capacity. As the stowing and picking processes are controlled by the retailer, we suppose that units of a SKU are picked and delivered to the customer according to the \textit{First In – First Out} (FIFO) principle, that is, the oldest SKUs are sold first.

When determining order quantities for the replenishment of a local fulfilment centre, the retailer needs to take into account several costs. If the number of units in the inventory exceeds customer demand, this generates costs for inventory holding denoted by $v$ per unit. In case of perishable SKUs, units may deteriorate beyond an acceptable level of quality at the end of a period. This is associated with a spoilage cost of $h$ per unit. On the other hand, unfulfilled customer demand leads to lost sales, whose costs are more difficult to determine (\citealp{Walter1975}, \citealp{Fisher1994}). These costs comprise short-term lost revenue and consequences of long term customer churn. Long-run objectives that impact expected future sales strongly affect the strategic service-level selection (\citealp{Anderson2006}). In e-grocery retailing, there is typically a strongly asymmetric cost structure due to the much increased risk of a complete order cancellation in case of a stock-out: the main convenience of online shopping, namely not having to visit a physical store, may then be outweighed by the inconvenience of the potential necessity of placing a second order, and generally of having to stay at home during the delivery time slot. Therefore, e-grocery retailers operate with very high service-level targets of 97-99\%. We follow previous literature on e-grocery retailing and derive cost parameters for lost sales from the strategic service level target of the e-grocery retailer addressing the trade-off between shortage costs and costs incurred by excess inventory (cf. \citealp{Ulrich2018}).

A particular advantage of the e-grocery retailing business case under consideration is the availability of uncensored demand data. This data allows to explicitly calculate the amount of lost sales for each period. Note that this is not possible in brick-and-mortar retailing as the observable demand in that case is limited to the inventory level.

\subsection{Dynamics of inventory process and resulting cost}
\label{sec:model-dynamics}

Given this problem description, the retailer in our business case is faced with an extensive number of SKUs, most with a shelf life exceeding one demand period, thus allowing the transfer of excess inventory at the end of any demand period to the following period (\citealp{kim2015optimal}). In such a multi-period setting, an SKU's replenishment order quantity for any period $t$ affects its inventory of the subsequent periods $t+1, t+2, \ldots$, leading to a dynamic inventory management problem. We denote the inventory\footnote{The inventory $i_t$ of an SKU available at time $t$ in general consists of units delivered at different replenishment instances. Let $\Tilde{i}_{t,j}$ be the number of units available in period $t$ but supplied $j$ periods ago. Then $i_t = \sum\limits_{j=0}^J \Tilde{i}_{t,j}$, where $J$ corresponds to the maximum shelf life of the SKU. Note, that therefore $i_t$ can be represented by a vector with elements $\Tilde{i}_{t,j}$. Note that \citet{siawsolit2021offsetting} follow a similar approach.} at the beginning of a demand period by $i_t$, $q_t$ corresponds to the quantity delivered at the beginning of period $t$, and demand is described by $d_t$. We assume that the replenishment order quantity $r_{t-\tau,t}$ to be delivered after a lead time $\tau$ and inducing the actually delivered quantity $q_t$ cannot be adjusted by the retailer after its specification in $t-\tau$.

We simplify the problem by assuming that the intra-period dynamics can be captured by a sequence of events occurring within period $t$ that affect the size of the inventory $i_t$. The progress resulting from these events is indicated by primes (see the visualisation of the timeline in Figure~\ref{fig:StructureDemandPeriod} and the numerical example in Appendix~\ref{sec:app_dynamics}). At the beginning of a demand period, there is a starting inventory $i_t$. The first thing happening in $t$ is the decision on the replenishment order $r_{t,t+\tau}$ affecting supply in the future period $t+\tau$. This decision has to be taken without information regarding the realisations of supply and demand in period $t$ that only become known later. After the order decision, the supply $q_t(r_{t-\tau,t})$ becomes known. This affects the inventory in the following form: $i'_t = i_t + q_t$. Next, demand $d_t$ becomes known.
Given the e-grocery business case, we assume that SKUs are picked from the inventory according to a FIFO principle. After taking the (satisfiable) demand out, the new inventory is written as $i''_t$. We assume that this inventory size marks the amount of the SKU under consideration that is affected by deterioration during period $t$; $z_t$ denotes the corresponding realisation of spoilage. Subtracting the spoilage yields the new inventory $i'''_t$ representing the inventory at the end of period $t$. This also gives the inventory at the beginning of the following period: $i_{t+1} = i'''_t$.

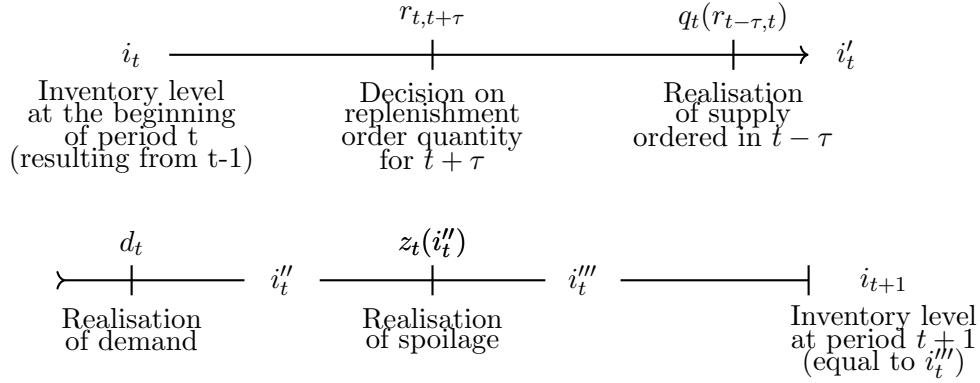
\begin{figure}[ht]
\begin{center}
\begin{tikzpicture}[scale=1]
\draw[->,thick] (0.5,0) -- (9,0);
\draw[>-,thick] (-1,-3) -- (1.5,-3);
\draw[-,thick] (2.5,-3) -- (5.5,-3);
\draw[-,thick] (6.5,-3) -- (9,-3);
\draw[-,thick] (4,-0.2) -- (4,0.2);
\draw[-,thick] (8,-0.2) -- (8,0.2);
\draw[-,thick] (0,-3.2) -- (0,-2.8);
\draw[-,thick] (4,-3.2) -- (4,-2.8);
\draw[-,thick] (9,-3.2) -- (9,-2.8);
\node [] at (0,0) {$i_t$};
\node [] at (0,-0.5) {Inventory level};
\node [] at (0,-0.8) {at the beginning};
\node [] at (0,-1.1) {of period t};
\node [] at (0,-1.4) {(resulting from t-1)};
\node [] at (8,0.5) {$q_t(r_{t-\tau, t})$};
\node [] at (8,-0.5) {Realisation};
\node [] at (8,-0.8) {of supply};
\node [] at (8,-1.1) {ordered in $t-\tau$};
\node [] at (9.5,0) {$i'_t$};
\node [] at (4,0.5) {$r_{t,t+\tau}$};
\node [] at (4,-0.5) {Decision on};
\node [] at (4,-0.8) {replenishment};
\node [] at (4,-1.1) {order quantity};
\node [] at (4,-1.4) {for $t+\tau$};
\node [] at (0,-2.5) {$d_t$};
\node [] at (0,-3.5) {Realisation};
\node [] at (0,-3.8) {of demand};
\node [] at (2,-3) {$i''_t$};
\node [] at (4,-2.5) {$z_t(i''_t)$};
\node [] at (4,-3.5) {Realisation};
\node [] at (4,-3.8) {of spoilage};
\node [] at (6,-3) {$i'''_t$};
\node [] at (4,-2.5) {$z_t(i''_t)$};
\node [] at (10,-3.0) {$i_{t+1}$};
\node [] at (10,-3.5) {Inventory level};
\node [] at (10,-3.8) {at period $t+1$};
\node [] at (10,-4.1) {(equal to $i'''_t$)};
\end{tikzpicture}
\end{center}
\caption{Sequence of events within one demand period.}
\label{fig:StructureDemandPeriod}
\end{figure}

In the lost sales case, we obtain the inter-period dependencies as:
\begin{equation}
    i_{t+1}=\max (i_t+q_t-d_t-z_t,0).
    \label{eq:IterativeScheme}
\end{equation}

At the end of each period $t$, costs occur depending on demand $d_t$ and the number of available units of the SKU $i_t+q_t$. In case of excess demand, costs for lost sales amount to $b$ per unit. However, excess inventory leads to either inventory costs of $v$, if the unit is still saleable in the subsequent period, or spoilage costs of $h$ per unit, since we assume that such units spoil at the end of the period under consideration. By assuming periodical replenishments, we can ignore fixed order costs in our model. Putting this together leads to the following costs obtained for period $t$:
\begin{equation*}
    C(r_{t-\tau,t}) = v \cdot \left(i_t+q_t-d_t-z_t\right)^+ +b \cdot \left(d_t-i_t-q_t\right)^+ + h \cdot z_t.
\end{equation*}

\subsection{Modelling uncertainty in the inventory management process}
\label{sec:model-uncertainty}

In general, retailers are faced with uncertainty in the number of units requested by customers in period $t$, rendering demand a random variable $D_t$ with cumulative distribution function (CDF) $F_{D_t}$. In addition, taking into account potential supply shortages, the quantity delivered by the supplier, $Q_t$, becomes stochastic and depends on the quantity ordered $r_{t-\tau,t}$, which it cannot exceed. If the relative supply shortage was known and constant, a retailer could simply add the percentage of known shortage to the specified replenishment order quantity to derive the target order quantity. However, supply shortages are neither constant nor known in retail practice, but rather follow an unknown probability distribution. In our model, we consider three different supply states $G_t$, namely complete delivery (state 1), a cancellation of the total delivery (state 2), and partial delivery (state 3), determining the relative proportion of supply $\delta_t$ in each demand period $t$. Since supply shortages often result from persistent problems in the supply chain, we model the sequence of supply states using a homogeneous Markov chain, specified by transition probabilities and its stationary distribution. In case of partial delivery, the proportion of units supplied is assumed to follow a Beta distribution with additional point masses on zero and one, respectively (\citealp{Ospina2012}).\footnote{See Appendix \ref{sec:appendix_supply} for technical details.}

To represent the more realistic case of uncertain shelf lives, which holds in particular for perishable SKUs as considered in our business case, we model the shelf life of an SKU in days using a discrete distribution estimated from historical data. As discussed in detail in Appendix \ref{sec:appendix_spoilage}, this distribution can be used to derive the conditional probability of a unit of the SKU deteriorating after a certain number of days, given it was still saleable in the period before. We denote the (stochastic) total number of deteriorated units at the end of period $t$ by $Z_t$.

Note that in case of a positive fixed lead time $\tau>0$, the inventory $I_t$ at the beginning of period $t$ is unknown at the decision instance $t-\tau$ when the order $r_{t-\tau,t}$ has to be placed. Rather it depends on $D_{t-\tau}, D_{t-\tau+1},\dots, D_{t-1} $ as well as the random yield and spoilage during the lead time. The distribution of inventory $I_t$ hence is a convolution of the corresponding probability distributions. 

Taking into account all stochastic variables affecting the inventory level leads to the following expected costs for period $t$ depending on the specified distributions:
\begin{equation}
    E[C(r_{t-\tau,t})] = v \cdot E\left(I_t+Q_t-D_t-Z_t\right)^++b \cdot E\left(D_t-I_t-Q_t\right)^+ + h \cdot E(Z_t).
    \label{eq:stochastic}
\end{equation}

Due to the lead time $\tau$, the replenishment order decision $r_{t-\tau,t}$ taken in period $t-\tau$ does not affect the cost in period $t-\tau$ but only the cost accumulated starting in period $t$. In this problem we aim to simultaneously consider consecutive periods affected by the replenishment order decision \citep{Alden1992}. Given the system dynamics in Equation~\ref{eq:IterativeScheme}, minimising the total expected costs over a planing horizon $T$
\begin{equation*}
\label{eq:expected_cost_total}
    \centering\min\limits_{r_{1,1+\tau}, \ldots, r_{T-\tau, T}}  \sum\limits_{t = 1}^{T-\tau}  E[C(r_{t,t+\tau})] \tag{SDLI}
\end{equation*}
yields the periodic review stochastic dynamic lost-sales inventory model with lead time (SDLI) under consideration in this paper. A comprehensive overview of lost sales inventory theory focusing on solution procedures for the different classes of lost sales models has been given by \citet{bijvank2011lost}. Accordingly, there is limited knowledge about optimal replenishment policies and no structure for an easy-to-understand optimal policy to be implemented in real-world applications. Similarly, \citet{boute2022deep} state that, in general, such models cannot be solved by exact methods due to the size of the state space of lost sales inventory models with lead time. Note that in our case the complicated nature of the convolution governing the development of the inventory level during the lead time even exacerbates this challenge \citep{halman2009fully}.

\subsection{Formulating the problem as a sequential decision process: interplay of information model and decision-making model}

Effectively supporting replenishment order decisions in e-grocery retailing does not only require to propose a solution procedure for deriving a replenishment policy for the inventory model (SDLI) but also to adequately incorporate relevant data available in the business environment under consideration. For example, historical data as well as contextual information improve the accuracy of the forecast on unknown probability distributions of future demand represented in (SDLI) (see e.g.\ \citealp{Bensoussan2007, Levi2007, Levina2010} for the use of data to solve certain inventory models). Given the structure and processes of the e-grocery business model, most data relevant to the operative replenishment decisions, such as the amount delivered to customers during the day or orders for future delivery placed by customers, become available on a day-to-day basis. Consequently, this requires to rerun a solution procedure for the inventory model (SDLI) each day and for each of the various SKUs based on the newly revealed information from the data collected during the previous day. Solving the inventory model in this way transforms (SDLI) into a decision model embedded in a sequential decision process. To represent the inventory management problem at hand as a sequential decision process, we follow the terminology and notation conventions proposed by \citep{Powell2019book}.

Core elements of a sequential decision process based on the data becoming available over time are an information model and, as already mentioned before, a decision model. The information model $\Omega$ covers exogenous information that becomes known over time and is stochastic for future periods. Realisations at the end of period $t$ are denoted by $\omega_t$ and cover information on realised demand, spoilage, and supply shortages in a given (or past) period(s) but also contextual information such as known demand for future periods and other features. Parts of these realisations are fed into prediction models that give estimates for the parameters of the underlying (non-stationary) probability distributions of the stochastic variables in the inventory model (SDLI).

\begin{figure}[ht]
\begin{center}
\scalebox{0.8}{
\begin{tikzpicture}[scale=1]
\draw[-,thick] (-0.5,0) -- (10,0);
\draw[-,thick] (-0.5,-1) -- (10,-1);
\draw[-,thick] (10,0) -- (10.5,-0.5);
\draw[-,thick] (10,-1) -- (10.5,-0.5);
\draw[-,thick] (0,0) -- (0.5,-0.5);
\draw[-,thick] (0,-1) -- (0.5,-0.5);
\draw[-,thick] (2,0) -- (2.5,-0.5);
\draw[-,thick] (2,-1) -- (2.5,-0.5);
\draw[-,thick] (4,0) -- (4.5,-0.5);
\draw[-,thick] (4,-1) -- (4.5,-0.5);
\draw[-,thick] (6,0) -- (6.5,-0.5);
\draw[-,thick] (6,-1) -- (6.5,-0.5);
\draw[-,thick] (8,0) -- (8.5,-0.5);
\draw[-,thick] (8,-1) -- (8.5,-0.5);
\node [] at (-2,-0.5) {realisation};

\draw[-,thick] (0.1,-2) -- (-0.4,-2.5);
\draw[-,thick] (0.1,-2) -- (0.6,-2.5);
\draw[-,thick] (0.1,-3) -- (-0.4,-2.5);
\draw[-,thick] (0.1,-3) -- (0.6,-2.5);
\draw[-,thick] (3.1,-2) -- (2.6,-2.5);
\draw[-,thick] (3.1,-2) -- (3.6,-2.5);
\draw[-,thick] (3.1,-3) -- (2.6,-2.5);
\draw[-,thick] (3.1,-3) -- (3.6,-2.5);
\draw[-,thick] (6.1,-2) -- (5.6,-2.5);
\draw[-,thick] (6.1,-2) -- (6.6,-2.5);
\draw[-,thick] (6.1,-3) -- (5.6,-2.5);
\draw[-,thick] (6.1,-3) -- (6.6,-2.5);
\node [] at (-2,-2.5) {prediction};

\node [] at (-3.75,-1.5) {\begin{sideways} information \end{sideways}};
\node [] at (-3.25,-1.5) {\begin{sideways} model \end{sideways}};
\node [] at (-3.75,-5.5) {\begin{sideways} decision \end{sideways}};
\node [] at (-3.25,-5.5) {\begin{sideways} model \end{sideways}};

\draw[-,thick] (0,-4) -- (1.5,-4);
\draw[-,thick] (0,-5) -- (1.5,-5);
\draw[-,thick] (0,-4) -- (0,-5);
\draw[-,thick] (1.5,-4) -- (1.5,-5);
\draw[-,thick] (3,-4) -- (4.5,-4);
\draw[-,thick] (3,-5) -- (4.5,-5);
\draw[-,thick] (3,-4) -- (3,-5);
\draw[-,thick] (4.5,-4) -- (4.5,-5);
\draw[-,thick] (6,-4) -- (7.5,-4);
\draw[-,thick] (6,-5) -- (7.5,-5);
\draw[-,thick] (6,-4) -- (6,-5);
\draw[-,thick] (7.5,-4) -- (7.5,-5);
\node [] at (-2,-4.5) {state};

\draw[thick] (0.75,-6.5) circle (15pt);
\draw[thick] (3.75,-6.5) circle (15pt);
\draw[thick] (6.75,-6.5) circle (15pt);
\node [] at (-2,-6.5) {decision};
\node [] at (0.75,-6.5) {${\scriptscriptstyle \text{SDLI}_1}$};
\node [] at (3.75,-6.5) {${\scriptscriptstyle \text{SDLI}_2}$};
\node [] at (6.75,-6.5) {${\scriptscriptstyle \text{SDLI}_3}$};

\draw[->,thick] (-0.5,-8) -- (10.5,-8);
\draw[-,thick] (0.75,-8.25) -- (0.75,-7.75);
\draw[-,thick] (3.75,-8.25) -- (3.75,-7.75);
\draw[-,thick] (6.75,-8.25) -- (6.75,-7.75);
\node [] at (0.75,-8.5) {$t = 1$};
\node [] at (3.75,-8.5) {$t = 2$};
\node [] at (6.75,-8.5) {$t = 3$};
\node [] at (11,-8) {time};
\draw[-,thick] (-1,0.5) -- (-1,-7.5);

\draw[->,thick] (0.1,-1) -- (0.1,-2);
\draw[->,thick] (3.1,-1) -- (3.1,-2);
\draw[->,thick] (6.1,-1) -- (6.1,-2);
\draw[->,thick] (0.1,-3) -- (0.1,-4);
\draw[->,thick] (3.1,-3) -- (3.1,-4);
\draw[->,thick] (6.1,-3) -- (6.1,-4);
\draw[->,thick] (0.75,-5) -- (0.75,-6);
\draw[->,thick] (3.75,-5) -- (3.75,-6);
\draw[->,thick] (6.75,-5) -- (6.75,-6);
\draw[->,thick] (1.5,-4.5) -- (3,-4.5);
\draw[->,thick] (4.5,-4.5) -- (6,-4.5);
\draw[->,thick] (7.5,-4.5) -- (9,-4.5);
\draw[-,thick] (1.25,-6.5) -- (3.1,-6.5);
\draw[->,thick] (3.1,-6.5) -- (3.1,-5);
\draw[-,thick] (4.25,-6.5) -- (6.1,-6.5);
\draw[->,thick] (6.1,-6.5) -- (6.1,-5);
\draw[-,thick] (7.25,-6.5) -- (9.1,-6.5);

\end{tikzpicture}}
\end{center}
\caption{Representation of our sequential decision process adapted from \citet{meisel2011anticipatory}.}
\label{fig:sqd}
\end{figure}
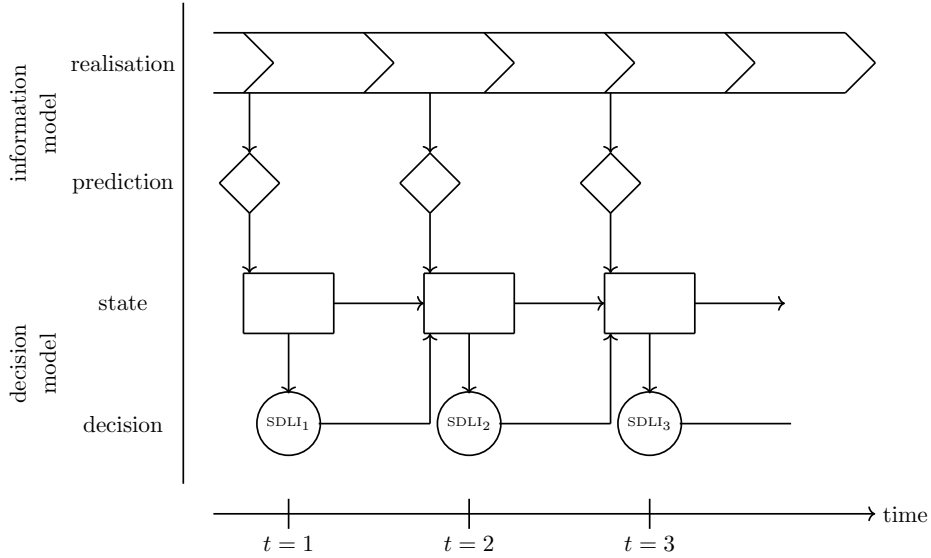

This description extends the notion and notation proposed by \citet{soeffker2022stochastic} in the context of a dynamic routing problem. Accordingly, the decision model uses a state $s_t$ and yields a decision $x_t$. Solving the decision model leads to the post-decision state $s_t^x$ and determines either an (expected) reward or an (expected) cost in period $t$. Given the state $s_t$, a decision $x_t$, and the realisation of the information model $\omega_t$, a transition function T gives the state in the following period $s_{t+1} = T(s_t, x_t, \omega_t)$. The decision $x_t$ is determined by a policy $\pi$: $x_t = X^{\pi}(s_t)$. The aim is to find an optimal policy $\pi^*$ that minimises/maximises the objective function given by either the sum of (expected) profits or (expected) costs over a planning horizon. A graphical representation of the sequential decision process for our inventory management business case is given in  Figure~\ref{fig:sqd}.

The model elements described in Sections~\ref{sec:model-dynamics} and \ref{sec:model-uncertainty} form the basis for the representation as a sequential decision process operating on the level of demand periods and using (SDLI) as a decision model. In our inventory management problem, the state $s_t$ comprises the inventory $i_t$ (with the corresponding vector $\Tilde{i}_t$ indicating the supply date as introduced in Section~\ref{sec:model-dynamics}), the supply state in the previous period $G_{t-1}$, the set of ordered (but not yet delivered) replenishment quantities $r_{t-\tau,t}, \ldots, r_{t-1, t+\tau-1}$ and the (estimated) probability distributions of customer demand, spoilage, and yield of future periods. In the post-decision state, the replenishment order quantity $r_{t,t+\tau}$ determined by the decision policy is known. Before the transition to the next state $s_{t+1}$ takes place, the stochastic variables, namely demand, supply, and spoilage, as well as contextual information, realise for period $t$, denoted by $\omega_t$, thus determining the resulting costs in this period. The other elements of the state variable $s_{t+1}$ are given by the supply state $G_{t}$, the set of replenishment order quantities that is augmented with $r_{t,t+\tau}$ by making an order decision in period $t$, and the estimated probability distributions of the various random variables affecting the future inventory level.

Note that the stochastic and dynamic decision model introduced here requires the inclusion of estimated probability distributions in $\omega_t$. Given the availability of comprehensive data sets in the e-grocery business environment this poses an opportunity and a challenge for descriptive and predictive analytics \citep{lepenioti2020prescriptive}. While a thorough discussion of these aspects in the light of supporting the e-grocery retailer's operative decision making is beyond the scope of this paper, we can refer to the analyses already carried out in detail by \citet{Ulrich2018} and \citet{Ulrich2021} for the demand part of our data set within the business case under consideration. We mention three crucial results of that work:

First, estimated demand distributions depend on the SKU under consideration, day of week, known demand, and other features; in particular, the demand distributions are not stationary. Second, \citet{Ulrich2018} compare a variety of statistical methods and machine learning approaches for the estimation of demand distributions. They establish that the estimation method yielding the best results varies with the SKU, day of week and other features. Therefore, third, to provide useful decision support in a business case, the choice of the estimation model has to be data-driven and automated. In \citet{Ulrich2021} a classification-based framework for such an automated procedure is proposed and evaluated. Because of the importance of data-driven estimation of the relevant probability distribution for the sequential decision problem in e-grocery, we amend the representation proposed by \citet{soeffker2022stochastic} accordingly (see Figure~\ref{fig:sqd}).

In our case, the determination of a replenishment order quantity $r_{t, t+\tau}$ corresponds to the \textit{decision} $x_t$ in the terminology from the literature. Instead of a reward, we consider the (expected) costs induced by the decision. We aim at finding a decision policy $\pi$, that is, a predefined reaction that minimises the (expected) costs induced by a decision $r_{t,t+\tau}$. Those costs are given by the expected immediate costs, i.e.\ costs for period $t+\tau$ where the decision affects the inventory management process, and the expected sum of future costs depending on the post decision state $s_t^r$, denoted as the \textit{value of the post decision state} $V(s_t^r)$.

For a given planning horizon $T$ and a policy $\pi$, we can formulate the sum of expected immediate and future costs to be minimised in period $t$ as follows:
\begin{equation}
    E\bigl[C_{t,t+\tau}^\pi \bigr] = E\bigl[ C(s_t, r^\pi_{t,t+\tau})\bigr] + \underbrace{E \bigl[ \sum\limits_{j = t + 1}^{T-\tau} C(s_j, r^\pi_{j,j+\tau} (s_j) \vert s_t, r^\pi_{t,t+\tau}) \bigr]}_{V(s_t^r)}
    \label{eq:sqd_costs}
\end{equation}
The expected immediate costs incurred by a policy $\pi$ leading to the decision in period $t-\tau$ can then be written by using the estimated probability distributions of the uncertain quantities (see Equation~\ref{eq:stochastic}); note that these estimates depend on the state $s_{t-\tau}$.

\subsection{A lookahead-based decision policy}
\label{sec:opt-policy}
For the solution of the decision model, we are looking for an order policy $\pi$ that, given a state $s_t$, minimises the expected cost (see Equation~\ref{eq:sqd_costs}). According to \citet{boute2022deep}, the best way to solve complex inventory control problems under uncertainty is to resort to approximate numerical methods. Specifically, the discussion in \citet{boute2022deep} focuses on Deep Reinforcement Learning (DRL) methods which rely on Deep Neural Networks (DNNs) for approximating the value of taking a certain decision in a certain state and/or for approximating an optimal policy function.  Following the classification of policies proposed by \citet{Powell2019book}, this approach can be categorised as a Value Function Approximation (VFA)-based policy and/or as a policy relying on policy function approximation (PFA). In our setting, we decided against relying on learning a VFA or a PFA since in the business case, this would mean having to train more than 100 ML models (e.g. DNNs). Furthermore, while \citet{boute2022deep} demonstrate that DRL approaches tend to work very well for stationary inventory management problems, our setting is highly  non-stationary. Given that the non-stationarity is only partly explainable by regular effects such as, seasonality, applying a DRL-based policy would require a frequent expensive re-training of the VFA / PFA models.

Instead of using a learned VFA model, we propose to approximate the value of a given order decision by a Monte Carlo simulation for a limited lookahead horizon $H$. Following \citet{Powell2019book}, the resulting policy can be characterised as a stochastic lookahead policy. A key advantage of this type of policy is that, using the terminology proposed by \citet{soeffker2022stochastic}, it makes internal use of the information model and thus naturally adapts to (even structural) changes in the information model without requiring to re-train an approximation model. As the policy relies on sampling, it can naturally be combined with advanced and context-dependent distributional forecasting approaches such as those proposed in \citet{Ulrich2018}.

When it comes to the lookahead horizon $H$, observe that the order decision taken at $t$ affects the objective function in period $t + \tau$ only, i.e.\ when the order is supposed to be delivered. Thus, we choose $H \geq \tau$. We denote the number of lookahead periods exceeding $\tau$ by $\nu$, that is, $H = \tau + \nu$. Let us first assume that $H=\tau$ (that is, $\nu=0$). In that case, for a given state $s_t$ and for a given replenishment order decision, induced by the policy $\pi$, $r^\pi_{t,t+\tau}$, we can approximate the expected cost $E(C(s_t, r^\pi_{t,t+\tau}))$ in period $t+\tau$ by simulating $N$ sample paths starting at period $t$ and ending at period $t+\tau$. For a sample path $n$, $C^n(s_t, r^\pi_{t,t+\tau})$ is obtained by simulating the state-transition logic described in Section \ref{sec:model-dynamics} from start state $s_t$ using the given decision $r^\pi_{t,t+\tau}$ and random samples from the distributions representing supply, demand, and spoilage in each of the simulated periods from $t$ to $t+\tau$. 
In this setting, the optimisation problem to be solved in each period $t$ reads as follows:

\begin{equation*} 
\label{eq:monte-carlo-approximation}
E\bigl(C(s_t, r^\pi_{t,t+\tau}\bigr) \approx \Bigl\{ \frac{1}{N}\sum\limits_{n=1}^N C^n(s_t, r^\pi_{t,t+\tau}) \Bigr\}.
\end{equation*}

If our lookahead horizon $H > \tau$, that is if $\nu > 0$, we extend the sample paths described above until the final period $t+H$ of the lookahead horizon. The motivation of doing so is to better capture the effect of the order decision on time periods beyond $t+\tau$. The costs in the lookahead periods after $t+\tau$ are not only affected by the decision $r_{t,t+\tau}$ to be taken in $t$, but also by the ``simulated'' decisions $r_{j,j+\tau}$ taken in periods $j$ with $t \leq j\leq t+\nu$ that are part of the lookahead. To account for the reduced precision of the forecast for future periods and to reflect the relative decrease in importance for the decision $r_{t,t+\tau}$, we weight the expected costs by the factor $\rho^{j-t}$ with $\rho\in (0,1)$ for periods $j \geq t$. Note that while the lookahead decisions $r_{j,j+\tau}$ for $j > t$ are not implemented, they are nonetheless part of the objective function of the decision model (SDLI). Hence the objective used for determining the lookahead policy reads as follows:

\begin{equation} 
\min_{r_{t,t+\tau}, \ldots, r_{t+\nu,t+\tau+\nu} } \Bigl\{ \frac{1}{N}\sum\limits_{n=1}^N \bigl( C^n(s_t, r^\pi_{t,t+\tau}) + \sum\limits_{j=t+1}^{t+\nu} \rho^{j-t} \cdot C^n(s^n_j, r^\pi_{j,j+\tau} (s^n_j) \vert s_t, r^\pi_{t,t+\tau}) \bigr) \Bigr\}
\label{eq:basic_lookahead_decision}
\end{equation}

Observe that the objective does not involve costs occurring in the periods before $t+\tau$, since they are not affected by the decisions involved in the lookahead. To determine the replenishment order quantity $r_{t,t+\tau}$ to be delivered in period $t+\tau$, we search for the quantity that minimises average costs over the sample paths as given in Equation~(\ref{eq:basic_lookahead_decision}) using a Nelder-Mead based numerical optimisation approach. In the following section, we will evaluate the proposed policy in a simulation study. In Section~\ref{sec:case}, we apply the policy to an e-grocery case study. In this case study, which is based on real-world data, we illustrate how the policy can be combined with distributional demand forecasts.

\section{Evaluation of the lookahead policy and the value of probabilistic information}
\label{sec:simulation-analysis}

The decision policy introduced above allows to explicitly consider the full uncertainty in the inventory management process by incorporating distributional information for the stochastic variables demand, spoilage, and supply shortage when determining replenishment order quantities. In practice, the underlying distributions for the stochastic variables need to be estimated, e.g.\ from historical data. However, the precision of the estimates of these probability distributions is highly dependent on the quality of the data available to the retailer. To avoid potential inaccuracies and allow for a comprehensive comparison between different policies, in this section, we rely on a simulation-based setting to evaluate the lookahead policy proposed above and to analyse the importance of incorporating probabilistic information when determining replenishment order quantities. Thus, we consider the simplified situation in which the retailer knows the probability distribution for each source of uncertainty (demand, spoilage, supply shortages), while we allow for non-stationarity and define the underlying distributions in accordance with a descriptive analysis of the data available in our business case.

Previous literature proposes the newsvendor model with its restrictive assumption of a shelf life of one period only to address the question on the optimal replenishment order quantity in case of stochastic customer demand. As a benchmark, we follow this myopic approach and thus ignore the possibility of transferring units to following periods as well as potential supply shortages. In the second step, we analytically calculate replenishment order quantities in the multi-period setting when referring to a deterministic setting using expected values of the relevant random variables as parameters of the decision model. Finally, we replace those expected values by probabilistic information (represented by distributions) to illustrate the possible advantages of our approach. The comparison to a myopic and a deterministic approach follows the suggestion by \citet{powell2009you}.

\subsection{Simulation setup: distributions, parameters, and data}
\label{sec:simulated-data}
We generate an experimental data set covering $T$ consecutive demand and supply periods for one example SKU. 
This data set provides information on demand, spoilage, and supply shortages.
Considering perishable SKUs with a shelf life of  one to multiple periods, we use a specific parameter vector for the data-generating process in the subsequent analyses.

\begin{figure}[ht]
\centering
\includegraphics[width=0.8\textwidth]{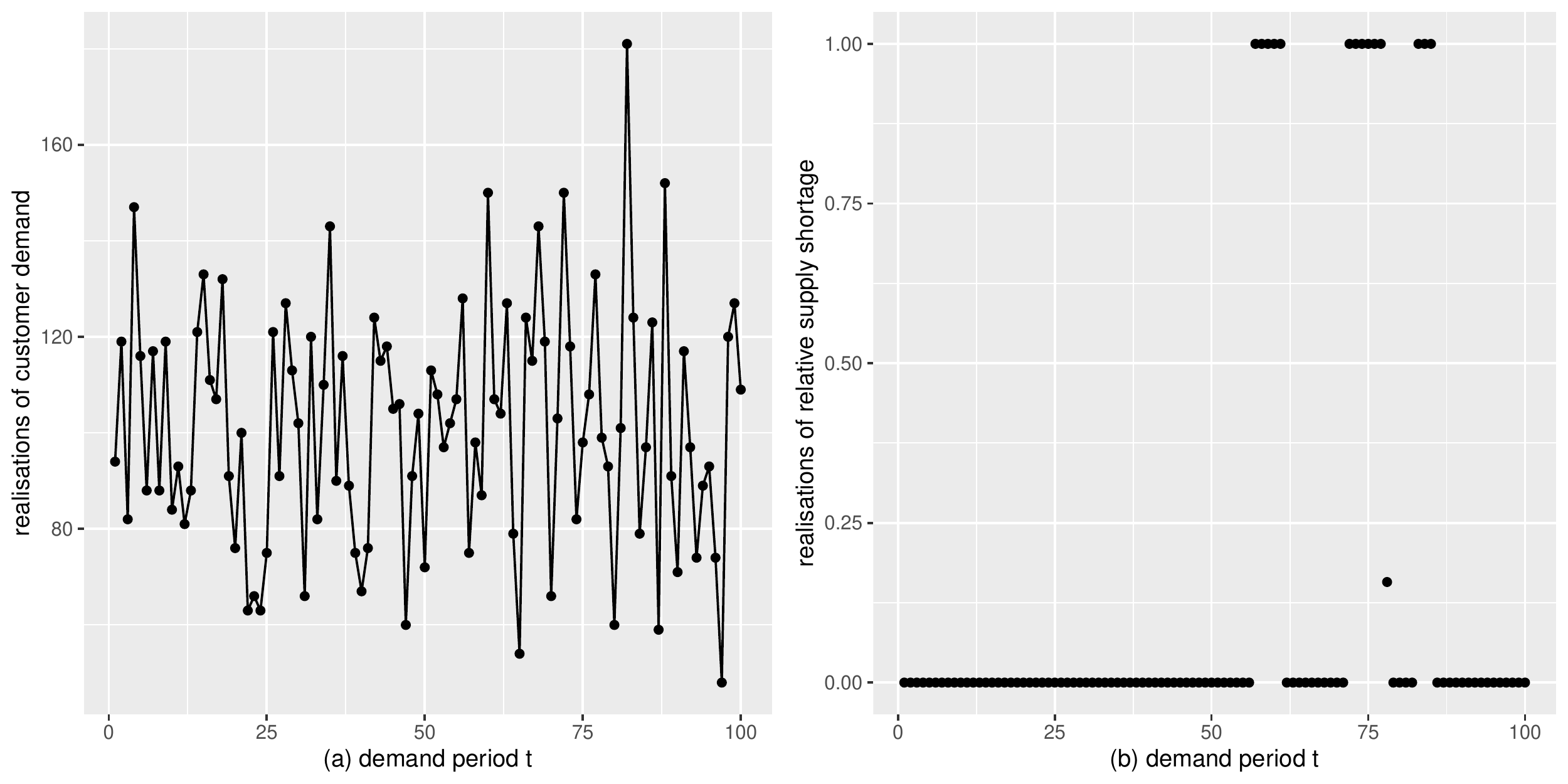}
\caption{Realisations of demand and shortage for demand period $t=1,\ldots,100$.} 
\label{fig:demandshortage}
\end{figure}

Based on the results on demand forecasting for e-grocery retailing by \citet{Ulrich2018}, we assume that the (uncensored) demand in period $t$ follows a negative binomial distribution with mean $\mu_t$ and size parameter $k_t$, i.e.\ $$ d_t \sim \text{NegBinom}(\mu_t, k_t).$$ We reparameterise the distribution in terms of its mean $\mu_t$ and variance $\sigma^2_t=\mu_t+\kappa_t$, with the relation $k_t=\mu_t/(\sigma^2_t-\mu_t)$. To allow for non-stationary demand, we draw the parameters of the demand distribution in each period as $\mu_t \sim Pois(\lambda_\mu)$ and $\kappa_t \sim Pois(\lambda_\kappa)$. For the subsequent analysis we assume $\lambda_\mu=100$ and $\lambda_\kappa=300$ and demand to be independent over the different periods, for simplicity avoiding more complex structures such as seasonality. An example realisation of simulated demand over 100 periods is shown in Figure \ref{fig:demandshortage} (a). We are able to generate data with similar patterns compared to those used in the case study (see Figure~\ref{figure_series_bananas}).

\begin{table}[ht]
    \centering
    \begin{tabular}{c|cccccc}
        $j$ & 1 & 2 & 3 & 4 & 5 & 6 \\
        \hline
        $f^{sl}(j)$ & 0.05 & 0.10 & 0.15 & 0.35 & 0.20 & 0.15 \\
    \end{tabular}
\vspace{0.5em}
    \caption{
Distribution of the shelf life in the simulated data set.}
    \label{tab:data_shelflife}
\end{table}

The shelf life of the SKU is generated from the distribution with probability mass function $f^{sl}(j)$, as shown in Table \ref{tab:data_shelflife}, with $j=1$ corresponding to the situation in which the unit deteriorates at the end of the delivery period (i.e.\ day 0). The mean shelf life implied by this distribution is three periods. The conditional probabilities of a unit deteriorating after exactly $j$ periods, given it was still saleable at the beginning of that period, are provided in Table~\ref{tab:spoilage} in the Appendix.

We assume there to be three ``states'' of delivery: complete delivery (state 1), complete shortage (state 2), and partial delivery (state 3), with the sequence of delivery states across demand periods governed by the Markov chain with transition probability matrix (TPM)
\begin{equation*}
\Theta=\left( \begin{array}{ccc}
    0.99 & 0.005 & 0.005 \\
    0.5 & 0.4 & 0.1 \\
    0.5 & 0.1 & 0.4 \\
\end{array}\right).
\end{equation*}
The associated stationary state distribution, also taken as the distribution for period $t=1$, is $\theta^* \approx (0.98, 0.01, 0.01)^t$. If the retailer is faced with partial supply shortage, i.e.\ if state 3 is active, then the realised relative amount of supply follows a beta distribution with shape parameters $\alpha=2$ and $\beta=3$, leading to an average relative shortage of 60\% in case of partial delivery and an overall average shortage of $\bar{\theta} = 1.57\%$. A corresponding example realisation of relative shortage for demand periods $t=1,\ldots,100$ is given in Figure \ref{fig:demandshortage} (b).

In accordance with the strategic environment given for the e-grocery retailer in our case study (see Section~\ref{sec:retailer-problem}), we assume the costs for one unit excess demand to be $b=5$, inventory costs to be $v=0.1$ per unit, and spoilage to generate costs of $h=1$ for the SKU considered. This relation between costs for excess inventory and shortages takes into account the high service-level target in e-grocery retailing. For the lookahead policy the absolute values of the cost parameters are not relevant, instead only the relation between these parameter values affects the solution determined by the model. A lead time of $\tau=3$ days is assumed to be required between the replenishment order decision and the delivery to the fulfilment centre of the retailer.

In the following, we provide analyses relying on a myopic approach, a deterministic approach and the lookahead policy introduced in Section~\ref{sec:opt-policy}. Our evaluation is based on $T = 5000$ simulated periods, which corresponds to more than 15 years of data in a business case. We use this many simulated data points to reduce the influence of noise in the comparison.

\subsection{First benchmark: myopic approach}

Previous literature on determining optimal replenishment order quantities commonly relies on the setting of the newsvendor model (see e.g.\ \citealp{Ulrich2018}). In this model it is assumed that each unit of a SKU can be sold for one demand period only, allowing to optimise the replenishment order quantity for each period individually. Each unit excess inventory leads to spoilage costs $h$, while each unit of lost sales leads to costs of $b$. The optimal order quantity can be obtained as the $b/(b+h)$--quantile of the (estimated) CDF of demand $F_D$ (\citealp{Silver1998}, \citealp{Zipkin2000}).\footnote{Note that $F_{D_{t+\tau}}$ corresponds to the forecast on demand taken in period $t$ for period $t+\tau$.} Under these myopic assumptions, the `optimal' replenishment order quantity $r_{t,t+\tau}^*$ can then be calculated as
\begin{equation}
    r^*_{t, t+\tau}=F_{D_{t+\tau}}^{-1}\left( \frac{b}{b+h} \right).
    \label{eq:newsvendor}
\end{equation}

For each period $t\in T$, we determine replenishment order quantities according to Equation~(\ref{eq:newsvendor}). Thus, we assume that the decision maker ignores the potential transfer of excess units at the end of a demand period and underestimates the starting inventory at the beginning of most demand periods, i.e.\ if there are units transferred. At the same time, the risk of supply shortages is ignored when determining replenishment order quantities. At the end of each period, realised inventory holding, spoilage, and lost sales generate costs according to the given cost parameters ($h$, $v$, $b$). We then calculate total costs over the time horizon considered $T$.

We obtain an average order quantity of 119.03, an average inventory quantity of 199.42 and an average amount of spoilage of 17.52 leading to average per period costs of 38.84. We are able to satisfy 99.72\% of total customer demand. This exceeds the strategic service level given by the retailer leading to higher costs than in the case the intended service level is exactly met, driven by additional costs for inventory holding as well as spoilage. This deviation from the desired service level is caused by ignoring the inter-period dependencies in the determination of replenishment order quantities within the newsvendor model, which is the classical inventory management model in case of stochastic customer demand. More complex models, which acknowledge the various stochastic factors in e-grocery retailing, are required to minimise the costs.

\subsection{Second benchmark: deterministic approach}

We now respect the dynamic relation of the inventory management problem under consideration, by taking into account that SKUs have an expected shelf life of multiple periods. More specifically, the expected shelf life is three periods with variation according to Table~\ref{tab:data_shelflife}. In addition, we consider the risk of supply shortages. Due to these two additional sources of uncertainty, the newsvendor model cannot be applied anymore. To still be able to derive analytical solutions, we rely on deterministic expected values for the stochastic variables impacting the inventory management process, namely demand, spoilage, and supply shortages. Thus, we still ignore the stochastic variation in supply shortages and the shelf life and, compared to the newsvendor approach, also ignore uncertainty in customer demand. In period $t$, we calculate the expected starting inventory for period $t+\tau$ denoted by $E(I_{t+\tau})$ according to the current inventory $i_t$, replenishment order quantities already determined $r_{t-\tau, t},\ldots, r_{t-1,t+\tau-1}$ as well as expected demand $\mu_{t},\ldots,\mu_{t+\tau-1}$ in the meantime, the average shelf life as well as the average amount of supply shortage. In case of a fixed relative supply shortage, the retailer could simply add this percentage to the replenishment order quantity to ensure the intended quantity to be delivered. The order quantity under this deterministic approach is then given by the difference between expected demand and the expected starting inventory in the period under consideration, divided by the average relative amount of units supplied:\footnote{Note that we consider only positive replenishment order quantities.}
\begin{equation}
    r_{t,t+\tau}^{*} = \max \left( \frac{E(D_{t+\tau}) - E(I_{t+\tau})}{1-\bar{\theta}}, 0 \right).
\end{equation}

Applying these point forecasts gives an average order quantity of 96.33 which is about 19\% lower compared to the newsvendor approach. At the same time, the average inventory level of 18.93 is more than 90\% lower. As a consequence, the amount of spoilage is also much reduced. This policy leads to a situation where 93.49\% of the total customer demand is fulfilled. This falls below the intended service level by the retailer. Thus, while costs for inventory holding and spoilage reduce compared to the newsvendor approach, lost sales occur more often and are responsible for a large share of total costs. However, total costs are slightly lower compared to the newsvendor approach. Respecting the inter-period dependencies in the inventory management framework by point forecasts on demand, shelf life and supply shortages reduces the average per-period costs by 8.5\% in total (see Table~\ref{tab:scenarios_results_1}).

\subsection{Evaluation of the lookahead policy}
\label{sec:simulated-EVIU}

In our setting with multiple sources of uncertainty (demand, supply, and spoilage), the use of point forecasts reduced average per-period costs for perishable SKUs compared to the more myopic newsvendor model, which addresses only the stochasticity of demand. However, the approach presented in the previous section results in more unfulfilled demand than intended by the strategic service level of the e-grocery retailer. Thus, in the following, we apply the lookahead policy introduced in Section~\ref{sec:opt-policy}, evaluating the policy in our simulation setting in detail. As the outcome of any of these policies is highly dependent on the business case, we provide a discussion on the sensitivity of our results with respect to the underlying parameter values, thereby generalising to other inventory management settings, in the online supplementary material.

We parameterise the policy based on a set of initial experiments, addressing the trade-off between computation time and stability of the simulation results. In particular, the retailer needs to determine order quantities for all SKUs in the assortment every day within a few hours, which limits the computing power and time available for single SKUs. Thus, based on a set of prior experiments, we use $N=1000$ sample paths (simulation runs), while considering $\nu=3$ additional periods with weighting factor $\rho=0.9$. In each period, the retailer determines the replenishment order quantity according to the lookahead policy and the (known) probability distributions for each source of uncertainty. At the end of a period, we again use realised inventory holding, spoilage, and lost sales to calculate average per period costs for the evaluation period.

Table~\ref{tab:scenarios_results_1} summarises the results of our analysis and compares them to those obtained under the newsvendor model and the deterministic approach. We find an average order quantity of 103.05, which is 7.0\% higher than when relying on expected values. At the same time, the average inventory level is more than threefold higher, while it is only about 30\% of the level under the newsvendor model. We observe that on average 98.47\% of customer demand is fulfilled, which is in the interval intended for the service level by the retailer. Average per-period costs when accounting for uncertainty in all three sources (demand, spoilage, supply) reduces by 52.0\% compared to the situation of point forecasts and even 56.1\% compared to the newsvendor model. This shows that our policy is able to provide more accurate replenishment order quantities compared to the myopic and deterministic approach.

\begin{table}[ht]
    \centering
    \begin{tabular}{l|llll|l}
            & $\varnothing$ order & $\varnothing$ inventory & $\varnothing$ amount & $\%$ fulfilled & $\varnothing$ costs \\ 
    setting & quantity          & level                 & of spoilage        &  demand        &  per period                 \\
    \hline
    newsvendor & 119.03 & 199.42 & 17.52 & 99.72\% & 38.84 \\
    \hline
    point forecasts & 96.33 & 18.93 & 0.99 & 93.49\% & 35.55 \\
    \hline
    lookahead policy & 103.05 & 59.16 & 3.53 & 98.47\% & 17.07\\
    \end{tabular}
    \vspace{0.5em}
\caption{Comparison of the lookahead policy to the myopic and deterministic approach.}
\label{tab:scenarios_results_1}
\end{table}

\subsection{Value of probabilistic information}
\label{sec:prob_inform}

While the application of the lookahead policy allows the retailer to account for uncertainty in the stochastic variables demand, supply, and spoilage in a multi-period setting where we assume underlying parameters for the probability distributions to be known, in practice, retailers need to adequately estimate these distribution from features such as historical data before they are able to make replenishment order decisions based on probabilistic information. To this end, data collection, data preparation, and data analysis require operational effort and costs for retailers, which needs to be taken into account. Thus, we now evaluate the benefit of applying distributional information for each of the different sources of uncertainty (limiting information on the other two sources to point forecasts). The results will give us insights into the value of probabilistic information. For each source of uncertainty, we consider two different information settings: (i) the retailer knows only the expected value of the uncertain quantity and (ii) the retailer knows the full probability distribution.

In the field of decision analysis, the improvement in expected performance resulting from using full distributional information is called \textit{expected value of including uncertainty} (EVIU), see e.g. \citet{Morgan1990} for a detailed description of EVIU and its relation to the value of information\footnote{A study addressing the value of information in the context of grocery retailing has been performed by \citet{siawsolit2021offsetting}.} in economics. In the context of stochastic programming, the same concept is typically referred to as \textit{value of the stochastic solution} (VSS), see e.g. \citet{Birge2011}. While most analyses regarding EVIU and VSS compare the consideration of distributions for \textit{all} stochastic variables to using no distributions at all, in the following investigation, we examine the value of considering distributions for each subset of the stochastic variables. In an actual business setting, such results can be compared to the costs incurred by the collection and processing of the data needed for obtaining the distributional information regarding the respective stochastic variable(s). In particular, this allows the retailer to decide for each source of uncertainty whether a probabilistic representation is cost-efficient. To give a comprehensive idea of the practical relevance of our approach, in the online supplementary material we additionally provide sensitivity analyses with respect to location and dispersion of the underlying probability distributions as well as the cost structure.

In our analysis, for each information scenario, the retailer optimises the replenishment order quantity in each demand period according to the information available (i.e.\ expected values or distributions). This allows us to estimate the EVIU, i.e.\ cost reductions gained from precise distributional information, for each source of uncertainty as well as for the whole model. Table~\ref{tab:scenarios_results_2} provides information on the different settings compared to probabilistic information for all sources of uncertainty, and additionally gives the savings compared to the newsvendor model as well as the setting of point forecasts. Including distributional information for demand only already leads to a comprehensive reduction in total costs relative to point forecasts (-51.6\%). To account for the variation in demand, the retailer here increases replenishment order quantities and holds a significantly higher safety stock. Therefore, the average inventory level and amount of spoilage increase more than threefold compared to the situation of point forecasts. However, because of the asymmetric cost structure, savings due to the increased service level exceed additional expenditures for spoilage and inventory holding. Improvements with respect to costs are obtained also when including only the shelf life's probability distribution, but with a much smaller effect --- -6.8\% total costs compared to point forecasts --- due to the low probability of spoilage within the first two sales periods.

In comparison to the deterministic benchmark (point forecasts only), considering a probability distribution exclusively for supply shortages decreases summary statistics on the average order quantity, inventory level and percentage of fulfilled demand. In particular, the average order quantity of 95.63 (see Table 3) is lower than the corresponding value of 96.33 under the deterministic approach (see Table 2). At the same time, the additional probabilistic information leads to an increase in average per-period costs. This is caused by the determination of replenishment order quantities according to Equation 6 in the benchmark case. Here, the retailer adds a fixed percentage to each order quantity to account for expected supply shortages. In total, this overcompensates stochastic supply shortages in most periods and leads to an increase in the average inventory level of 13\%. As a consequence of the higher inventory level generated by less information on supply shortages in the deterministic case, holding and spoilage costs increase. However, the increased inventory level also has a positive effect, indicated by the comparatively higher percentage of fulfilled demand in the deterministic case; it unintentionally reduces lost sales and the corresponding cost incurred by disregarding the variation in demand in the deterministic setting. Due to cost parameter asymmetry, therefore, average total costs are lower for the less well informed decision maker.

\begin{table}[ht]
    \centering
    \scalebox{0.75}{
    \begin{tabular}{l|llll|lll}
            & $\varnothing$ order & $\varnothing$ inven- & $\varnothing$ amount & $\%$ fulfilled & $\varnothing$ costs  & cost change & cost change \\ 
    setting & quantity & tory level & of spoilage & demand & per period & rel.\ to NV & rel.\ to PF \\
    \hline
    distributional infor- & 103.86 & 60.72 & 3.37 & 98.5\% & 17.20 & $-55.7\%$ & $-51.6\%$ \\
    mation for demand &  &  &  &  &  &  & \\
    \hline
    dist.\ info.\ & 96.92 & 20.11 & 1.06 & 94.0\% & 33.13 & $-14.7\%$ & $-6.8\%$ \\
    for shelf life &  &  &  &  &  &  & \\
    \hline
    dist.\ info.\ for & 95.63 & 16.76 & 0.88 & 92.9\% & 38.08 & $-2.0\%$ & $+7.1\%$ \\
    supply shortages &  &  &  &  &  &  & \\
    \hline
    dist.\ info.\ for each & 103.05 & 59.16 & 3.53 & 98.5\% & 17.07 & $-56.1\%$ & $-52.0\%$ \\
    uncertainty source &  &  &  &  &  &  & \\
    \end{tabular}}
    \vspace{0.5em}
\caption{Analysis of the expected value of including uncertainty (NV: newsvendor model; PF: point forecasts).}
\label{tab:scenarios_results_2}
\end{table}

Table~\ref{tab:app_scenarios_results} in the Appendix provides additional results on combinations where we apply distributional information for two sources of uncertainty while in the third source relying on the expected value. We find that the value of including uncertainty varies between the different model components, while also the sequence of including distributional information matters. For example, we find that including the probability distribution for supply is only beneficial when the retailer also accounts for uncertainty in demand.

The above simulation-based analysis indicates how retailers can reduce total costs when using full probability distributions instead of expected values for each source of uncertainty, when underlying probability distributions are known. However, it is to be expected that the results strongly depend on the exact specification of the distributions of the random variables associated with demand, supply, and shelf life (and of course also on the cost structure). We hence provide a sensitivity analysis in the supplementary material, where we gradually change the parameters used for each source of uncertainty and derive the resulting costs. We observe that the benefit of incorporating the demand distribution increases when its variance increases, which can be explained by the asymmetric cost structure. Incorporating information on the shelf life distribution is most beneficial for distributions with a high variance or a small mean (corresponding to a high risk of spoilage in early periods). The relevance of incorporating probabilistic information on potential supply shortages depends not only on the associated risk, but also on the persistence of the corresponding process (i.e.\  whether shortages tend to occur in several consecutive periods). Regarding the cost structure, we find that when assuming a constant relationship between inventory costs and spoilage costs, then potential savings increase with a higher cost asymmetry (due to lost sales).

\section{Case study}
\label{sec:case}

The simulation-based analysis in the previous section demonstrates the importance of respecting inter-period dependencies and basing the decision on replenishment order quantities on probabilistic information instead of expected values. However, in practice, the underlying distributions need to be estimated from historical data, typically making use of features (covariates) to arrive at time-varying predictive distributions. Thus, we now use the setting and data from a European e-grocery retailer to illustrate the analysis process in a situation where we need to integrate both parameter estimation and optimisation. In the following, we first give an overview of the data set available to us, followed by the case-specific tuning of the lookahead policy introduced in Section~\ref{sec:opt-policy}. We compare the results under our proposed policy to those obtained when applying the decision rule as currently implemented by the retailer.

\subsection{Data}
\label{sec:case-data}
\begin{figure}[ht]
\centering
\includegraphics[width=0.75\textwidth]{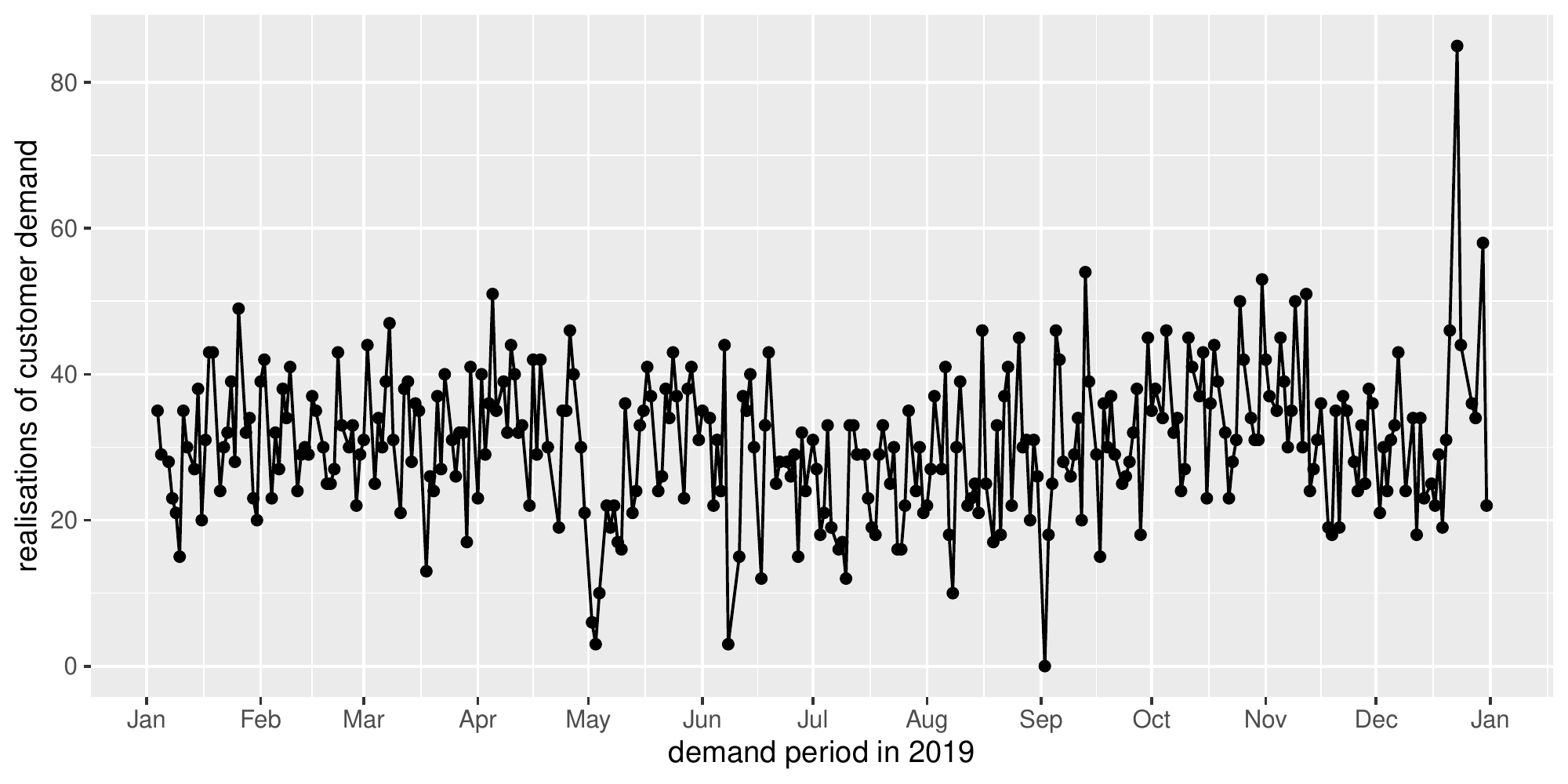}
\caption{Realised demand for the SKU mushrooms in 2019.} 
\label{figure_series_bananas}
\end{figure}

The data set on the attended home delivery service provided by the e-grocery retailer covers demand periods of six different local fulfilment centres from January 2019 to December 2019, i.e.\ before the beginning of the Covid-19 pandemic. One observation here equals one demand period $t$, i.e.\ one day of delivery. We consider four SKUs within the category fruits and vegetables, namely mushrooms, grapes, organic bananas, and lettuce. For illustration, Figure \ref{figure_series_bananas} displays the demand for the SKU mushrooms in 2019 for one selected fulfilment centre. We find recurring peaks on Mondays, but do not observe any notable trend or seasonality. The data set includes features to be used for the demand forecast as well as the (uncensored) realised demand in this period. For a more detailed description, we refer to \citet{Ulrich2018}.

For the perishable SKUs analysed, the number of sales periods before spoilage is not defined by best-before-dates, but may depend on non-constant prior supply chain attributes, such as the weather or the country of origin. Due to missing best-before-dates, our data set includes a parameter for the expected number of sales periods for each SKU, predefined by the retailer. The expected number of sales periods for lettuce, as an example, equals one demand period, i.e.\ it is assumed that excess inventory cannot be sold in the following demand period and thus generates spoilage. In addition, the data set includes information on the quantity ordered, the quantity delivered by the regional distribution centre, and the number of units deteriorated in a certain demand period.

\subsection{Application of our lookahead policy}
\label{sec:optimisation-policy}

At the beginning of each demand period, the current inventory level, the supply state of the last period, and previous replenishment order quantities for future demand periods within the lead time are known. From historical data, we estimate the distribution of demand in future periods, the transition probability matrix (TPM) to make predictions with respect to possible supply states, and the distribution of shelf lives --- details are provided below. The data set provided by the e-grocery retailer allows us to make use of uncensored demand data. Realised supply shortages, however, are observed only for the quantity that was requested, and hence not for other possible order quantities. Here, we consider the relative supply shortage and apply this value to the order quantity determined under the lookahead policy. In addition, data related to spoilage depends on previous replenishment order decisions, as these quantities affect the level of inventory and, therefore, the amount of spoilage observed in the data set. Thus, we simulate spoilage according to the underlying probability distribution observed in the data set for the corresponding month and fulfilment centre.

In our simulation-based analysis in Section~\ref{sec:prob_inform}, we found that it is important to incorporate full probabilistic information for customer demand, while the additional value of incorporating uncertainty in spoilage and supply was rather small. Thus, in this case study, we endeavour to build precise probabilistic predictions of demand, but refrain from using complex statistical modelling of supply shortages and shelf life (e.g.\ using features).

The demand forecast is obtained via regression modelling, considering the features \textit{ID of the fulfilment centre}, \textit{weekday},  \textit{price}, \textit{marketing activities}, \textit{known demand}, and \textit{median demand of the previous month}. The marketing activities were included only for the SKU grapes, as for the others there was no marketing campaign in the demand periods analysed. Based on the good performance of distributional regression methods in situations with very high service-level targets in \citet{Ulrich2018}, we apply generalised additive models for location, scale and shape (GAMLSS) for demand forecasting, assuming a negative binomial distribution for the response.

Supply shortages are assumed to be governed by a 3-state Markov chain with state transition probabilities estimated from historical data. For the state associated with partial supply, the parameters of the beta distribution are estimated based on historical partial supply shortage. The implied stationary distributions   of supply states for the SKU mushrooms are given in Table \ref{tab:supply_shortage}. Results show that about 98.2\% of all replenishment orders are fully served by the national and local distribution centres. Incomplete supplies and complete supply shortages both occur with a probability of about 0.9\%.

\begin{figure}[ht]
\centering
\includegraphics[width=0.5\textwidth]{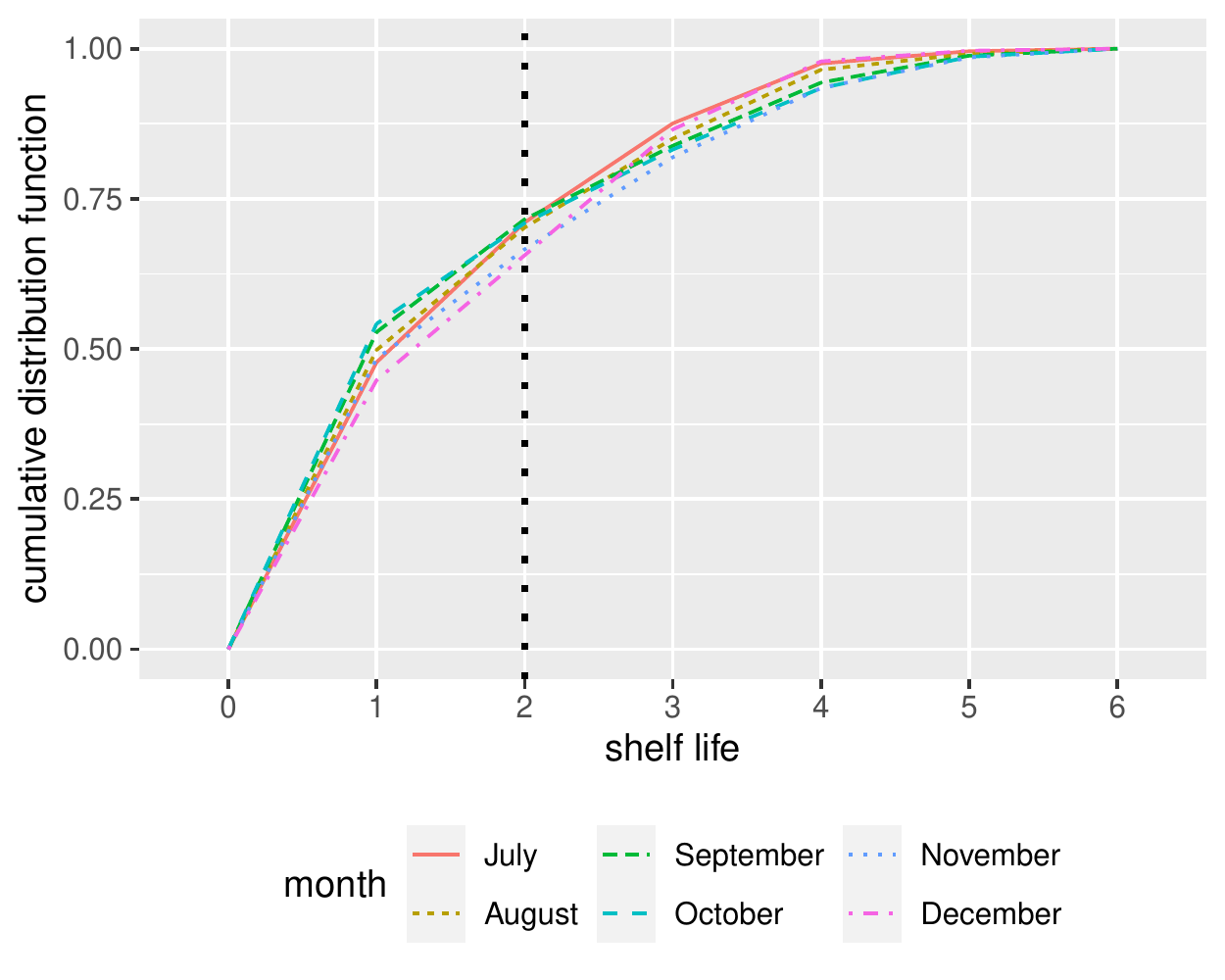}
\caption{Estimated CDF for the shelf life of the SKU mushrooms for July-December, aggregated over all fulfilment centres.}
\label{fig:CDF_ShelfLife}
\end{figure}

To derive the stochastic distribution of the shelf life for a given SKU, we consider the number of units deteriorated at the end of a certain period, for which we calculate the supply date under the assumption of the FIFO principle based on historical data. This allows us to derive the relative frequencies of shelf lives within the data set. Figure \ref{fig:CDF_ShelfLife} illustrates the estimated CDF of the shelf life for the SKU mushrooms. We find only slight differences between months, implying a low level of seasonality in the shelf life of this SKU.\footnote{In the analysis, we assume a maximum shelf life of six days and add the remaining probability to day 1-6 on a proportional basis.} While about 30\% of the units have a shelf life larger than two days, every other unit already deteriorates after the first day.

For each source of uncertainty and each SKU, we use the previous six months of data to estimate the associated probability distributions and incorporate them into the lookahead policy for an evaluation period of one month. For example, we train on data from January to June 2019 to forecast demand, spoilage, and supply shortages in July 2019. Due to the limited number of demand periods during six months, we aggregate historical data on spoilage and supply shortage over the fulfilment centres to ensure stable estimations.

\begin{table}[ht]
    \centering
    \begin{tabular}{c|cccccc|c}
        supply state & Jul & Aug & Sep & Oct & Nov & Dec & $\varnothing$ \\
        \hline
        full delivery & 0.9878 & 0.9812 & 0.9793 & 0.9780 & 0.9773 & 0.9869 & 0.9818 \\
        no delivery & 0.0012 & 0.0082 & 0.0105 & 0.0116 & 0.0113 & 0.0103 & 0.0089 \\
        partial delivery & 0.0110 & 0.0106 & 0.0102 & 0.0104 & 0.0113 & 0.0028 & 0.0094 \\
    \end{tabular}
    \vspace{0.5em}
    \caption{Stationary distribution of supply shortage states for the SKU mushrooms for July-December 2019, aggregated over all fulfilment centres.}
        \label{tab:supply_shortage}
\end{table}

\subsection{Benchmark policy of the retailer}
\label{sec:benchmark-policy}

As a benchmark to our lookahead policy, we replicate the current replenishment order decision process according to the guidelines of the e-grocery retailer considered. For each SKU, the retailer operates with an inventory level target that equals the sum of the expected mean demand\footnote{We use the same forecast on mean demand as in the lookahead policy (see Section \ref{sec:optimisation-policy}).} and a fixed percentage of the expected mean demand as safety stock. The safety stock depends on historic realised mean demand and the expected number of sales periods. SKUs with a low mean demand and a low number of expected sales periods are complemented with a low safety stock, e.g.\ 30\% of the mean demand for lettuce, whereas SKUs with a high mean demand and a high number of expected sales periods are provided with higher safety stocks, e.g.\ 70\% of the mean demand for grapes (cf.\ Table \ref{tab:benchmark_parameters}).

For each SKU, the expected number of sales periods is specified by a fixed parameter, e.g.\ one sales period for the SKU lettuce (cf.\ Table \ref{tab:benchmark_parameters}). Therefore, the retailer does not consider any variation in the shelf life of the SKU. The comparison between the assumed fixed shelf life of two days for the SKU mushrooms and the CDF in Figure~\ref{fig:CDF_ShelfLife} provides evidence for potential cost reductions by incorporating stochastic spoilage instead of a fixed sales period into the inventory management process.

The e-grocery retailer further assumes that the quantity delivered equals the quantity ordered, i.e.\ the yield rate equals 100\%. As a consequence, the risk of random yield is neglected by the retailer and does not impact the replenishment order decision.

\begin{table}[ht]
    \centering
    \begin{tabular}{l|cccc}
        SKU & safety stock & sales periods & yield rate\\
        \hline
        mushrooms & 50\%  & 2 & 100\%\\
        grapes & 70\% & 3 & 100\% \\
        bananas & 50\% & 2 & 100\%\\
        lettuce & 30\% & 1 & 100\% \\
    \end{tabular}
    \vspace{0.5em}
    
    \caption{Safety stock as a share of mean demand, the expected number of sales periods and the expected yield percentage for the SKUs analysed in our case study.}
    \label{tab:benchmark_parameters}
\end{table}

\subsection{Details on the implementation}
\label{sec:case-implementation}

We estimate the CDFs of shelf lives and the TPMs for the supply states based on a rolling window procedure covering six months of previous data updated in every month. Consequently, we evaluate the lookahead policy according to an out-of-sample data set, e.g.\ we train on data for January to June and evaluate throughout July. This enables a comparison between the suggested lookahead policy and the benchmark for six consecutive months from July to December 2019. For both approaches we assume that the inventory is empty at the 1\ts{st} of July, i.e.\ at the beginning of our evaluation period. Due to the lead time of $\tau=3$, we consider the replenishment order quantities before the 4\ts{th} of July as given and identical for both policies. For each demand period, we conduct two steps. First, the next replenishment order quantity is determined according to the underlying policy, and second, the period is evaluated using the business case data to calculate the costs that the respective policy would have generated. For Sundays and bank holidays, i.e.\ when there was no service, we set the replenishment order quantity to zero.

As our demand data is uncensored, it does not depend on the inventory level (see the characteristics on e-grocery retailing in Section~\ref{sec:e-grocery}). Therefore, we are able to evaluate the lookahead policy according to the \textit{true} demand which in particular is not limited by the demand fulfilled under the policy of the retailer. However, as our replenishment order quantity for a given period may deviate from the quantity actually ordered by the retailer for the corresponding day, we make use of the information on the relative amount of incompletely supplied replenishment order quantities in the data of the retailer. We transfer this information to our order quantities, i.e.\ if there was full supply (or full shortage), we also assume full supply (or full shortage) for a different quantity. The number of units deteriorating depends on the composition of the inventory with corresponding date of supply. Since the inventory in our model again deviates from the inventory given by the retailer's data, we use simulations to determine which number of units would have been deteriorated if the retailer had followed the policy. Specifically, as introduced above, we assume spoilage to follow a binomial distribution with the probability parameter estimated from historical data and the number of units with a given supply date. To determine the amount of spoilage in the evaluation, we make use of the underlying probability distribution which is used as the forecast in the lookahead policy for the following month.\footnote{As an example, we calculate a CDF of shelf life for a SKU based on spoilage between February and July, and use this distribution to (1) calculate spoilage embedded in the lookahead policy of the replenishment order decision for August and (2) evaluate resulting spoilage in July.} For each demand period and supply date we generate a random number from a uniform distribution on (0,1). Applying this value to the inverse CDF of the shelf life gives the number of units deteriorated. Using the same random number in the evaluation of both approaches ensures that a larger inventory on a given day with identical supply date leads to a larger number of units deteriorated and vice versa.

We calculate total realised costs by considering costs for inventory holding, spoilage, and demand shortages using the replenishment order quantity determined by our lookahead policy and the benchmark. Cost parameters are given as introduced in Chapter \ref{sec:simulation-analysis} by $v=0.1$ for one unit in the inventory, $h=1$ for each unit deteriorated, and $b=5$ for one unit of lost sales. Evaluating both policies for each SKU and fulfilment centre enables us to monitor the resulting inventory at the end of a period, the number of units deteriorated, lost sales, and resulting total costs for each demand period within the evaluation period.

\subsection{Results}
\label{sec:case-results}
We evaluate four SKUs within six fulfilment centres. Due to missing data for the SKU lettuce in two fulfilment centres, we are able to evaluate 22 SKU/fulfilment centre combinations in total. Table \ref{tab:case_results} illustrates relative changes in the resulting average costs, i.e.\ relative savings, when using our lookahead policy instead of the benchmark approach. Overall, we find substantial cost reductions of 6.2\% to 23.7\% for all four SKUs. As our data set allows us to evaluate only six months of data, results vary considerably across the different SKU/fulfilment centre combinations, and for 4 out of the 24 combinations we do in fact see an increase in costs. In particular, for the combinations where we obtain substantially higher costs under the lookahead policy (grapes and lettuce in fulfilment centre 4), we find that realised demand is considerably lower than the forecast. This underlines the importance of a high quality of probability distribution estimation before the integration in an optimisation framework. It should also be noted that the cost parameters used in the lookahead policy may differ from the cost structure implicitly embedded in the benchmark policy. However, our results in the sensitivity analysis (cf.\ the Supplementary Material) show that using probabilistic information is superior across different values of the cost parameter for lost sales $b$.

\begin{table}[ht]
    \centering
    \begin{tabular}{c|rrrr}
         ID & mushrooms & grapes & bananas & lettuce \\
         \hline
        1 & $-41.3\%$ & $-10.6\%$ & $-26.5\%$ & $-20.8\%$ \\
        2 & $-11.1\%$ & $-26.6\%$ & $-37.9\%$ & NA \\
        3 & $-29.2\%$ & $-31.6\%$ & $-34.3\%$ & $-26.2\%$ \\
        4 & $-11.3\%$ & $+24.4\%$ & $+1.0\%$ & $+44.6\%$ \\
        5 & $-13.0\%$ & $+5.6\%$ & $-14.5\%$ & NA\\
        6 & $-26.6\%$ & $-7.5\%$ & $-6.1\%$ & $-6.6\%$ \\
        \hline
        $\varnothing$ & $-23.7\%$ & $-8.8\%$ & $-20.7\%$ & $-6.2\%$ \\
    \end{tabular}
    \vspace{0.5em}
        
    \caption{Average change in the relative costs when using our lookahead approach compared to the benchmark approach, for each fulfilment centre and SKU.}
    \label{tab:case_results}
\end{table}

Our simulation study in Chapter 4 suggests that retailers are already able to reduce costs substantially even when accounting only for demand uncertainty. Therefore, we further compare average costs when using the lookahead policy incorporating only information on the demand distribution with the benchmark policy for the SKU mushrooms and every fulfilment centre (Table \ref{tab:mushrooms_detail}). We find that using only the demand distribution reduces average costs over all fulfilment centres by 22.9\%, whereas additionally including distributional information on the shelf life and supply shortages leads to a further cost reduction of only 1.1\%. These findings corroborate the results from the simulation study, indicating that the demand distribution is the main source of uncertainty and the most relevant information to incorporate in the replenishment order decision. 

\begin{table}[ht]
    \centering
    \scalebox{0.8}{
    \begin{tabular}{l|rrrrrr|r}
        fulfilment centre ID & 1 & 2 & 3 & 4 & 5 & 6 & $\varnothing$ \\
        \hline
        LAP with demand distribution & $-41.2\%$ & $-18.3\%$ & $-29.9\%$ & $+3.9\%$ & $-20.3\%$ & $-20.2\%$ & $-22.9\%$ \\ 
        vs.\ benchmark & & & & & & & \\ \hline
        LAP with all distributions vs.\ & $-0.1\%$ & $+8.7\%$ & $+1.0\%$ & $-14.6\%$ & $+9.2\%$ & $-8.0\%$ & $-1.1\%$ \\ LAP with demand distribution & & & & & & & \\ \hline
        LAP with all distributions & $-41.3\%$ & $-11.1\%$ & $-29.2\%$ & $-11.3\%$ & $-13.0\%$ & $-26.6\%$ & $-23.7\%$ \\ vs.\ benchmark & & & & & & & \\
    \end{tabular}}
    \vspace{0.5em}
    
    \caption{Relative changes in average costs under different lookahead policies (LAPs) for the SKU mushrooms.}
    \label{tab:mushrooms_detail}
\end{table}

\begin{figure}[ht]
\centering
\includegraphics[width=1\textwidth]{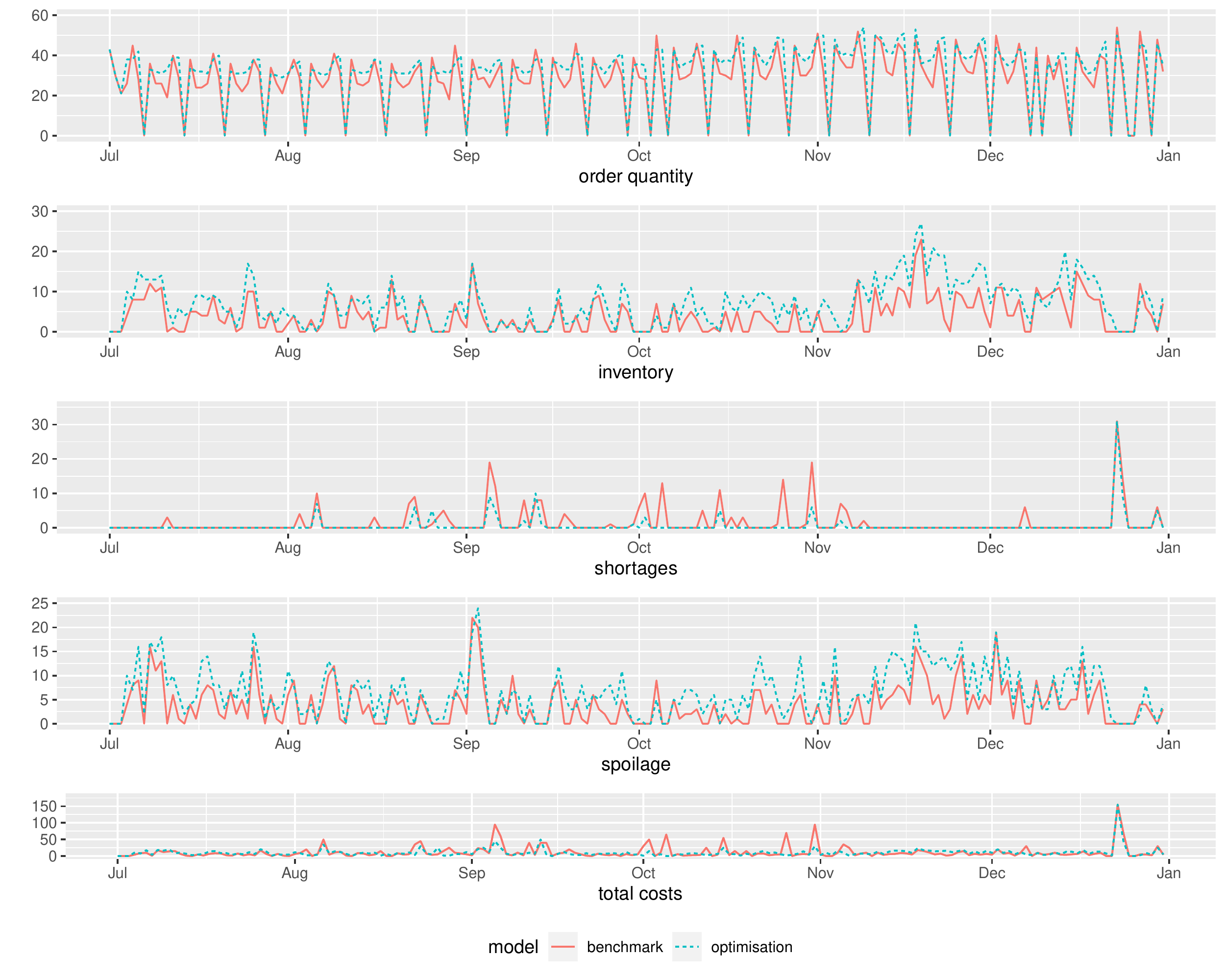}
\caption{Order quantities, inventory, shortages, spoilage, and total costs for the SKU mushrooms from fulfilment centre 4.}
\label{results_231010_1928831}
\end{figure}

Figure \ref{results_231010_1928831} shows detailed results for the SKU mushrooms in fulfilment centre 4, displaying the order quantities, inventory level, shortages, spoilage, and total realised costs for the lookahead policy (blue dotted line) and the benchmark model (red solid line). In total, there are 154 demand periods with a positive demand forecast for this SKU/fulfilment centre combination. In most demand periods (108 out 154), the order quantity obtained under our lookahead policy is larger than under the benchmark policy. The average replenishment order quantity under the lookahead policy is 31.87, compared to 28.23 under the benchmark model, with the main differences occurring midweek. As a consequence, the average inventory under the lookahead policy (6.94) is also lager than under the benchmark model (4.00). The difference in the inventory level between both approaches increases in the second half of the evaluation period, namely in October, November, and December.

For both approaches, the inventory level at the end of a period and the number of deteriorated units is highly correlated (correlation coefficient $\geq 0.7$), and as a consequence our lookahead policy yields a higher spoilage. In contrast, the number of lost-sales occurrences due to an unavailability is larger under the benchmark model (37 periods with an average number of 1.45 lost sales) than under our lookahead policy (16 periods with 0.58 lost sales on average).

\begin{figure}[ht]
\centering
\includegraphics[width=0.75\textwidth]{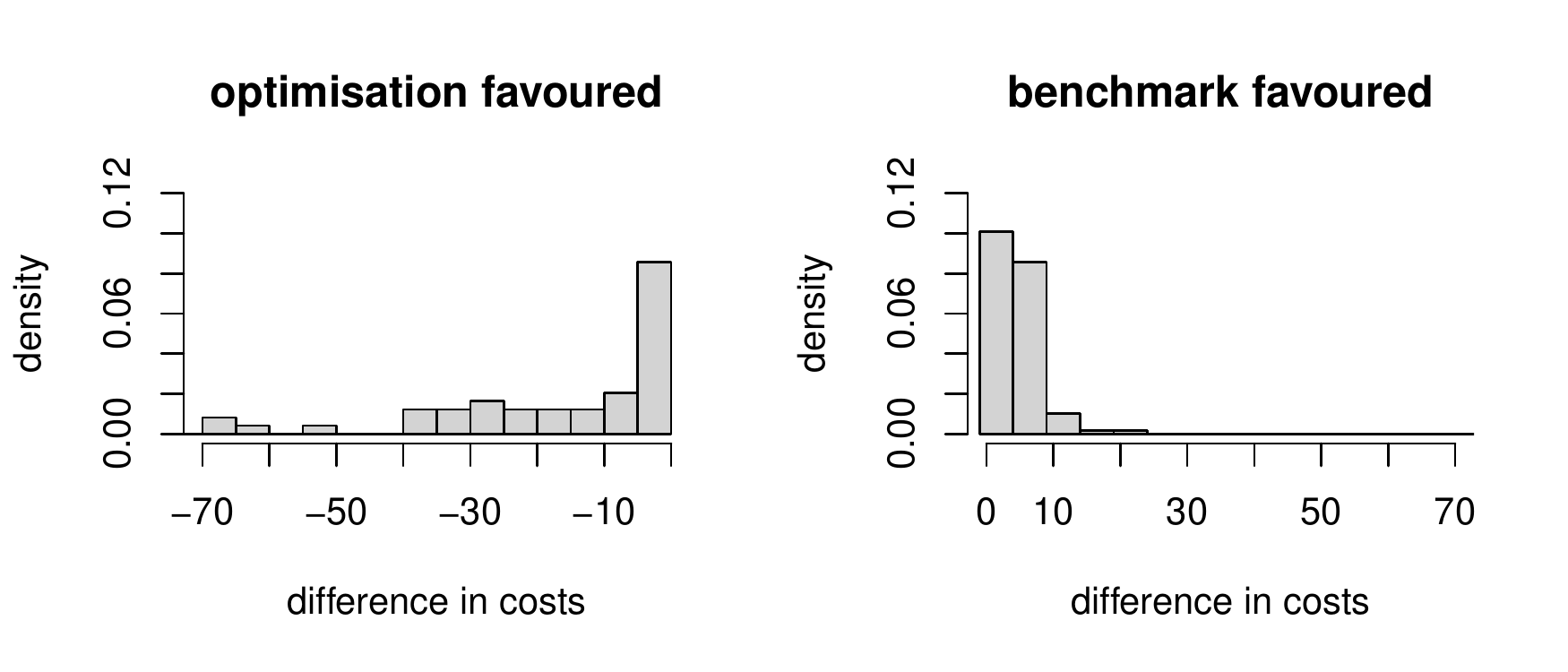}
\caption{Difference in costs between benchmark and lookahead policy. The positive sign corresponds to higher costs under the benchmark approach.} \label{fig:costs_diff}
\end{figure}

Lost sales are more costly for retailers due to long-term consequences such as unsatisfied customers switching to another company. In our setting with the specific assumptions made for the cost parameters, the higher safety stock under the lookahead policy induces lower average costs over the full evaluation period. The asymmetric cost structure leads to the interesting result that we find higher costs under the lookahead policy in about 70\% of the demand periods, yet the average overall costs are lower by about 11.3\% (see Table \ref{tab:case_results}). To illustrate this phenomenon, Figure \ref{fig:costs_diff} displays histograms of the single-period cost differences between the two approaches. The right panel covers the 119 demand periods with higher costs under our lookahead policy (positive sign), with an average difference of 4.56. In contrast, the average difference in costs in the 49 periods where the costs are higher under the benchmark policy (left panel) is $-15.98$, hence much higher (in absolute value).

In summary, we find that when fully accounting for the uncertainties in inventory management, the asymmetric cost structure in e-grocery retailing leads to higher average replenishment order quantities. While resulting costs under the lookahead policy are slightly increased for the majority of periods due to higher inventory levels and spoilage, the minimisation of lost sales yields an overall reduction in costs for the retailer compared to the benchmark policy.

\section{Conclusion}

In this paper, we propose a stochastic lookahead policy embedded in a data-driven sequential decision process for determining replenishment order quantities in e-grocery retailing. We aim at investigating to what extent this approach allows a retailer to improve the inventory management process when faced with multiple sources of non-stationary uncertainty, namely stochastic customer demand, shelf lives, and supply shortages, a lead time of multiple days and demand that is lost if not served. To this purpose, we represent the determination of replenishment order quantities as solutions of a dynamic stochastic period-review inventory model with lost sales and an expected-cost objective function. In real-world applications, the probability distributions of the inventory level at the beginning of a period and its marginals, such as distributions of demand, spoilage, and supply shortage, are typically unknown and hence need to be estimated. The periodically updated estimates for these distributions form the states in the sequential decision process; the inventory model plays the role of a decision model (see Figure~\ref{fig:sqd}). The analysis of data provided by the business partner was carried out in previous studies using descriptive and predictive methods \citep{Ulrich2018,Ulrich2021}; the findings are applied in the numerical analyses of this paper. The literature stresses the difficulty of finding an optimal replenishment policy for decision models like the one discussed here. We therefore propose a stochastic lookahead policy that allows us to integrate probabilistic forecasts for the underlying probability distributions into the optimisation process in a dynamic multi-period framework. We thereby demonstrate the feasibility of the integration of the different components of the data-driven sequential decision process (analytics and statistics, modelling and optimisation). In addition, the framework enables us to gain insights into the value of probabilistic information in our environment, not least in order to find some guidance for designing an adequate decision model. Finally, we show that such a framework is applicable to a real-world business environment of e-grocery retailing, potentially to the benefit of the retailer.

Evaluating an experimental data set generated in accordance with data provided by our business partner, we can show that our approach yields a replenishment policy that reduces the corresponding inventory management costs compared to the frequently applied newsvendor model. In addition, we analyse the value of explicitly exploiting probabilistic information instead of relying on point forecasts (expected values) in our replenishment decisions. Our results demonstrate that incorporating the full distributional information for all sources of uncertainty can lead to substantial cost reductions (with the amount of savings of course depending on the specific situation). The importance of including distributional information tends to increase with higher asymmetry in cost parameters (i.e.\ very low or very high service-level targets), as commonly found in e-grocery retailing. Regarding the different sources of uncertainty, the simulation results indicate that the benefit of integrating the probability distributions instead of expected values when determining replenishment order quantities is highest for customer demand. In contrast, the additional contribution of modelling shelf lives and supply shortages by probability distributions here turns out to be marginal but highly dependent on the structure of the underlying probability distributions (see the analyses in the online supplementary material). Finally, in a case study based on a comprehensive data set provided by a European e-grocery retailer we demonstrate the practical applicability of our approach by comparing the order policy under our approach to a policy used by this company in practice. Considering four different SKUs, we obtain cost savings between 6\% and 25\% when averaging over six fulfilment centres. From a managerial perspective, the simulation-based analyses as well as the case study suggest that using prescriptive analytics relying on modern computational methods to exploit the considerable amount of data available in e-grocery retailing is beneficial for retailers. In particular, it has the potential to outperform simple parametric inventory management policies designed by experienced human experts as well as myopic policies such as those based on the simple newsvendor model and deterministic approaches based on expected values. In addition to explicitly accounting for all sources of uncertainty, a key advantage of our lookahead policy over simple parametric policies is that it naturally adapts to a changing environment (e.g.\ induced by dynamic market developments), structural shocks (e.g.\ the Covid-pandemic), and regime shifts due to strategic changes (e.g.\ an increased focus on sustainability). Furthermore, it easily allows an adaption to the business cases of other companies. Specifically, our sensitivity analyses already provide a generalisation to other cases and present results to be expected in different settings.

Future work can extend this approach in several directions. In our case the retailer's management formulates the relevant goals for the e-grocery unit via a strategic service level requirement. Based on a set of initial experiments, in our numerical analyses, we assumed a certain relation between the lost-sales cost parameter and the other per-unit costs. In the newsvendor model there is a well-known analytical relationship between the cost ratio and the service level under the optimal solution; in our multi-period decision model solved by the lookahead policy, such a relationship must be explored numerically. At the same time, in order to carry out the numerical analysis, estimates on the underlying probability distributions are required. As demonstrated by \citet{Ulrich2018}, the quality of an estimation method is highly case-specific. To address this issue, statistical theory proposes to consider the loss function resulting from the decision problem in question as a benchmark. Hence, in order to choose an adequate estimation method, the lost-sales cost parameter needs to be known, i.e.\ the optimisation of the decision model and the choice of the estimation approach should be carried out simultaneously. This issue is also addressed in the outlook of \citet{boute2022deep}, who suggest to integrate parameter estimation in the optimisation of replenishment order quantities, and is directly related to very recent discussion on `predict-and-optimize' (see e.g.\ \citealp{Elmachtoub2021,vanderschueren2022predict}). In addition, our results demonstrate a relatively small impact of the stochastic variables supply and spoilage. However, relying the probabilistic forecasts on contextual data such as e.g.\ weather conditions might lead to an increased explanatory power of the underlying variation. While we already allow for non-stationary parameters but consider the same type of distribution for customer demand, \citet{Ulrich2021} show that the most promising model is highly case-dependent and may vary over time for the same SKU. Thus, future research could account for this finding and extend the flexibility for the integration of uncertainty in customer demand in our inventory management framework. Previous literature also discusses the case of advanced demand information in the setting of grocery retailing (see e.g.\ \citealp{siawsolit2021offsetting}) and the benefit of a potential discount rewarded to customers who place their order in advance. Such aspects could also be included in our stochastic inventory model. Due to the flexibility of our approach, extensions like these can be easily integrated into our inventory modelling framework.

\bibliographystyle{apalike} 
\bibliography{library}

\newpage
\appendix
\section{Appendix}

\subsection{Numerical example on the dynamics of the inventory system}
\label{sec:app_dynamics}

We consider an exemplary demand period $t$. We assume that the SKU under consideration has a shelf life of six periods. In the beginning, 50 units are kept in the inventory, 10 of these units were delivered by the supplier in period $t-2$ and 40 units in period $t-1$. This leads to the inventory vector $\Tilde{i}_t = (0,40,10,0,0,0)$. We consider the replenishment order quantity $r_{t-\tau,t} = 60$, while we assume a relative shortage of 20\% for period $t$. This leads to a delivered quantity of $q_t = 48$ and the adjusted inventory vector $\Tilde{i}'_t = (48, 40, 10, 0, 0, 0)$. We assume demand is $d_t = 46$. According to the FIFO principle, we primarily sell units from earlier periods, i.e.\ $t-1$ and $t-2$. This gives the inventory vector $\Tilde{i}''_t = (48, 4, 0, 0, 0, 0)$. Finally, we assume that 2 out of 4 units from period $t-1$ deteriorate while 12 out of 48 units delivered in period $t$ deteriorate. This gives $z_t = 14$ and the final inventory vector $\Tilde{i}'''_t = (36, 2, 0, 0, 0, 0)$ to be transferred to $\Tilde{i}_{t+1} = (0, 36, 2, 0, 0, 0)$. The transition is summarised in Table~\ref{tab:dynamics_example}.

\begin{table}[ht]
    \centering
    \begin{tabular}{c|cccc}
        Action &  & $q_t = 48$ & $d_t = 36$ & $z_t = 14$ \\
        \hline
        Resulting inventory vector & $\Tilde{i}_t$ & $\Tilde{i}'_t$ & $\Tilde{i}''_t$ & $\Tilde{i}'''_t$ \\
        \hline
        Entry & & & & \\
        $t,0$ & 0 & 48 & 48 & 36 \\
        $t,1$ & 40 & 40 & 4 & 2 \\
        $t,2$ & 10 & 10 & 0 & 0 \\
        $t,3$ & 0 & 0 & 0 & 0 \\
        $t,4$ & 0 & 0 & 0 & 0 \\
        $t,5$ & 0 & 0 & 0 & 0 \\
    \end{tabular}
    \caption{Inventory vectors according to the example used for the illustration of the dynamics given in Appendix~\ref{sec:app_dynamics}.}
    \label{tab:dynamics_example}
\end{table}

\subsection{State distribution of supply shortage}
\label{sec:appendix_supply}

Let $\delta_t$ denote the proportion of the ordered quantity $r_t$ that is actually supplied, such that $1-\delta_t$ is the relative supply shortage. The homogeneous Markov chain determining the sequence of supply states $G_1,\ldots,G_T$ is specified by the transition probabilities $\theta_{i,j}=\Pr(G_t=j \vert G_{t-1}=i)$, $i,j \in\{0,1,2\}$, $t \geq 2$. The state distribution in the first period $t=1$ is assumed to be given by the Markov chain's stationary distribution, $\boldsymbol{\theta}^*=\left( \Pr(G_t=1), \Pr(G_t=2), \Pr(G_t=3) \right)$. The proportion of units supplied, $\delta_t$, is then determined as follows:
\begin{equation}
\delta_t=
\begin{cases}
    1 & \text{if} \,\,\, G_t=1 \\
    0 & \text{if} \,\,\, G_t=2 \\
    \text{Beta}(\alpha,\beta) & \text{if} \,\,\, G_t=3.
\end{cases}
\end{equation}
In case of partial delivery, the beta distribution assumed for the proportion of units delivered implies a mean supply rate of $\alpha/(\alpha+\beta)$ --- the sum $\alpha+\beta$ constitutes a precision parameter. Across all three states, the proportion of units supplied follows a beta distribution with additional point masses on zero and one and a stationary mean of ${\theta}_1^* + {\theta}_3^* \cdot \alpha/(\alpha+\beta)$.

\subsection{Calculation of conditional probabilities for spoilage}
\label{sec:appendix_spoilage}
The conditional probability $p_j$ that a given unit deteriorates after $j$ periods is given by
\begin{equation}
    p_j=
    \begin{cases}
        f^{sl}(j) & j=0; \\
        \frac{f^{sl}(j)}{1-F^{sl}(j-1)} & j>0,
    \end{cases}
\end{equation}
where $f^{sl}$ is the probability function of shelf life learned from data and $F^{sl}$ the corresponding CDF. The inventory is represented by a vector $\Tilde{i}_{t,j}$ as introduced in Section~\ref{sec:app_dynamics} to keep track of the different delivery periods of units in stock. Given the probability $p_j$, the number of units from a set of $\Tilde{i}_{t,j}$ units with same supply date and a shelf life of $j$ deteriorating at a given day can be modelled by a binomial distribution with the parameters $\Tilde{i}_{t,j}$ and $p_j$. Hence the total number of deteriorated units at the end of period $t$, $Z(i_t)$, results from the joint distribution of these $J$ binomial distributions for the elements of inventory vector $\Tilde{i}_{t,j}$, each with individual parameters. All remaining units are transferred to the following period.

\begin{table}[ht]
    \centering
    \label{tab:spoilage}
    \begin{tabular}{c|c|c|c|c|c|c|c|c|c}
        Demand period $j$ & 1 & 2 & 3 & 4 & 5 & 6 \\
        \hline
        Spoilage probability $p_j$ & 0.050 & 0.105 & 0.176 & 0.500 & 0.571 & 1.000
    \end{tabular}
        \caption{Conditional probability of spoilage $p_j$ at the end of a given demand period $j$ in the simulated data set.}
\end{table}

\newpage
\subsection{Additional results on the effect of probabilistic information}

\begin{table}[ht]
    \centering
    \scalebox{0.65}{
    \begin{tabular}{c|ccc|cccc|c}
         & \multicolumn{3}{c}{distributional information on} & average & average & average & average & average\\
         & demand & shelf life & supply & order quantity & inventory level & amount of spoilage & service level & per period costs\\
         \hline
         Scenario 1 & & & & 96.33 & 18.93 & 0.99 & 93.49\% & 35.55 \\
         Scenario 2 & & & x & 95.63 & 16.76 & 0.88 & 92.91\% & 38.08 \\
         Scenario 3 & & x & & 96.92 & 20.11 & 1.06 & 94.00\% & 33.13 \\
         Scenario 4 & & x & x & 96.27 & 17.97 & 0.94 & 93.48\% & 35.44 \\
         Scenario 5 & x & & & 103.86 & 60.72 & 3.37 & 98.46\% & 17.20 \\
         Scenario 6 & x & & x & 104.09 & 62.86 & 3.51 & 98.55\% & 17.07 \\
         Scenario 7 & x & x & & 103.79 & 59.84 & 3.33 & 98.44\% & 17.16 \\
         Scenario 8 & x & x & x & 103.86 & 60.47 & 3.36 & 98.47\% & 17.07 \\
    \end{tabular}}
    \caption{Statistics on the average order quantity, inventory level, amount of spoilage, service levels, and per period costs for all scenarios.}
    \label{tab:app_scenarios_results}
\end{table}

\newpage

\section*{Supplementary material}

\subsection*{Sensitivity analysis for the results in Chapter 4}
\label{app:sensitivity}
For each of the three sources of uncertainty, we compare the results obtained under the deterministic approach (denoted as Scenario 1) and under full probabilistic information (denoted as Scenario 8) in the same simulation setting. For an overview on all scenarios see Table~\ref{tab:app_scenarios_results} in the Appendix. For the analyses we again refer to $T=5000$ simulation periods and the parametrisation of the lookahead policy as introduced in Section~\ref{sec:prob_inform}.

\begin{figure}[ht]
\centering
\includegraphics[width=0.6\textwidth]{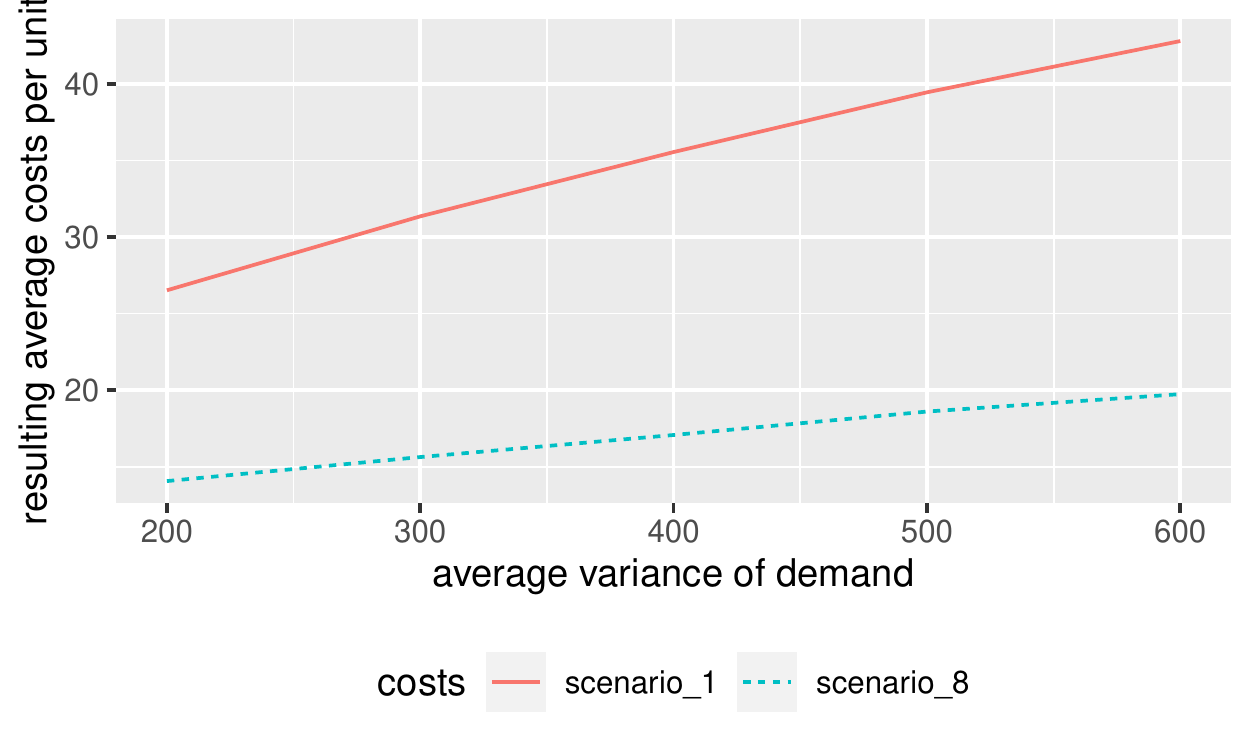}
\caption{Resulting per-period costs depending on the variance of demand.} \label{fig:sensitivity_demand}
\end{figure}

We start by adjusting the variance of demand. As our costs are asymmetric, with lost sales more expensive than inventory and spoilage, we expect the benefit of incorporating the demand distribution to increase when its variance increases. We still allow for non-stationary demand but vary the parameter in the variance-generating Poisson distribution, $\lambda_\kappa\in \{100,200,400,500\}$, holding $\lambda_\mu=100$ constant. Figure~\ref{fig:sensitivity_demand} shows that the average costs substantially increase when using expected values only (Scenario 1), while the per-period costs only slightly increase when incorporating full distributional information for all sources of uncertainty (Scenario 8). As expected, the importance of incorporating information on the demand distribution thus increases with its variance. 

The sensitivity of the results with respect to the shelf-life distribution is analysed in two different ways. First, we consider two settings ($f_1^{sl}$ and $f_2^{sl}$) with the same mean shelf life (three periods) but different variability. Second, we analyse two shelf-life distributions ($f_3^{sl}$ and $f_4^{sl}$) with the same relatively small variance but different mean shelf lives. The distributions are provided in Table \ref{tab:data_shelflife_sensitivity}. Here $f^{sl}_1$ corresponds to an SKU with a small variation in the shelf life, with 70\% of the units deteriorating one day after the expected shelf life at the latest, and each unit being saleable for 2--5 periods. In contrast, $f^{sl}_2$ represents a heavy-tailed distribution where both short shelf lives (one period) and longer ones (six periods) are quite likely. Distribution $f^{sl}_3$ corresponds to a situation where 80\% of the units spoil within the first two demand periods and a mean shelf life of two periods, while with $f^{sl}_4$ the average shelf life is five periods.

\begin{table}[ht]
    \centering
    \begin{tabular}{c|cccccc}
        j & 1 & 2 & 3 & 4 & 5 & 6 \\
        \hline
        $f^{sl}(j)$ & 0.05 & 0.10 & 0.15 & 0.35 & 0.20 & 0.15 \\
        \hline
        $f_1^{sl}(j)$ & 0 & 0.1 & 0.25 & 0.7 & 0.05 & 0 \\
        \hline
        $f_2^{sl}(j)$ & 0.2 & 0.05 & 0.05 & 0.25 & 0.15 & 0.3 \\
        \hline
        $f_3^{sl}(j)$ & 0.4 & 0.4 & 0.075 & 0.075 & 0.025 & 0.025 \\
        \hline
        $f_4^{sl}(j)$ & 0.025 & 0.025 & 0.075 & 0.075 & 0.4 & 0.4 \\
    \end{tabular}
    \caption{Distributions of shelf life in the sensitivity analysis.}
    \label{tab:data_shelflife_sensitivity}
\end{table}

\begin{table}[ht]
    \centering
    \scalebox{0.8}{
    \begin{tabular}{cl|l|l|l}
         & & & Probabilistic information & Full probabilistic \\
        set & distribution of shelf life & Deterministic approach & on shelf life & information \\
        \hline
        $f^{sl}$ & baseline & $35.55$ & $33.13$ ($-6.8\%$) & $17.07$ ($-52.0\%$) \\
        $f^{sl}_1$ & small variance & $33.88$ & $33.83$ ($-0.0\%$) & $14.75$ ($-56.4\%$) \\
        $f^{sl}_2$ & high variance & $40.01$ & $34.05$ ($-14.9\%$) & $22.38$ ($-44.1\%$) \\
        $f^{sl}_3$ & small mean & $44.52$ & $37.28$ ($-16.3\%$) & $27.28$ ($-38.7\%$) \\
        $f^{sl}_4$ & high mean & $34.81$ & $33.28$ ($-4.4\%$)  & $15.63$ ($-55.1\%$) \\
    \end{tabular}}
    \caption{Comparison of resulting average per-period costs for Scenarios 1, 3, and 8 depending on the distribution of shelf life. Relative savings compared to the deterministic approach (Scenario 1) in brackets.}
    \label{tab:sensitivity_shelflife}
\end{table}

Table \ref{tab:sensitivity_shelflife} provides an overview on the resulting average per-period costs under Scenario 1 (using expected values only), Scenario 3 (using distributional information for shelf life only) and Scenario 8 (using full distributional information). In the baseline setting according to the data set introduced in Section~\ref{sec:simulated-data}, the distribution is nearly symmetric around the mean shelf life of three periods, with a small risk of spoilage within the first two periods. In this setting, cost reductions of around 52\% could be achieved when incorporating full distributional information, while the reduction in Scenario 3 is limited to 6.8\%. If the risk of a very early spoilage is low, as caused by a small variance ($f^{sl}_1$) or a high mean ($f^{sl}_4$), similar cost reductions are achieved. In contrast, incorporating distributional information for shelf life only (Scenario 3) is more beneficial for distributions with a high variance ($f^{sl}_2$) or a small mean ($f^{sl}_3$), corresponding to a high risk of spoilage in early periods. At the same time, due to increased total costs, reductions achieved when incorporating probability distributions for all sources of uncertainty (Scenario 8) are smaller than under the baseline distribution.

Next we analyse the sensitivity of the results with respect to supply shortages, considering four different transition probability matrices regarding the change of supply states while holding the parameters of the beta distribution in case of partial supply shortage constant. The first matrix corresponds to a situation where the retailer is a bit more often faced with complete shortage than under the baseline scenario, with rare switches to partial or full shortage:

\begin{equation*}
\Theta_1=\left( \begin{array}{ccc}
    0.95 & 0.01 & 0.04 \\
    0.3 & 0.2 & 0.5 \\
    0.3 & 0.5 & 0.2 \\
\end{array}\right), \quad \theta^*=(0.857, 0.062, 0.081)^t
\end{equation*}
In the second setting, with 
\begin{equation*}
\Theta_2=\left( \begin{array}{ccc}
    0.8 & 0.199 & 0.001 \\
    0.199 & 0.8 & 0.001 \\
    0.495 & 0.495 & 0.001 \\
\end{array}\right), \quad \theta^*=(0.4995, 0.4995, 0.001)^t
\end{equation*}
partial supply in the next period occurs with probability 0.001 regardless of the current state. The other two states, full supply and full shortage, occur equally often.
Within the last two settings, the stationary probabilities are identical across all three states.
The difference between these two settings lies in the state persistency, with $\Theta_3$ corresponding to higher and $\Theta_4$ to lower persistence:
\begin{equation*}
\Theta_3=\left( \begin{array}{ccc}
    0.9 & 0.05 & 0.05 \\
    0.05 & 0.9 & 0.05 \\
    0.05 & 0.05 & 0.9 \\
\end{array}\right), \quad \pi^*=(1/3, 1/3, 1/3)^t,
\end{equation*}

\begin{equation*}
\Theta_4=\left( \begin{array}{ccc}
    1/3 & 1/3 & 1/3 \\
    1/3 & 1/3 & 1/3 \\
    1/3 & 1/3 & 1/3 \\
\end{array}\right), \quad \pi^*=(1/3, 1/3, 1/3)^t.
\end{equation*}

\begin{table}[ht]
    \centering
    \begin{tabular}{c|l|l|l}
         & & Probabilistic information & Full probabilistic \\
        set & Deterministic approach & on supply shortages & information \\
        \hline
        Baseline & $35.55$ & $38.08$ ($+7.1\%$) & $17.07$ ($-52.0\%$) \\
        $\Theta_1$ & $56.97$ & $42.96$ ($-24.6\%$) & $42.96$ ($-24.6\%$) \\
        $\Theta_2$ & $195.15$ & $188.48$ ($-3.4\%$) & $182.81$ ($-6.3\%$) \\
        $\Theta_3$ & $164.28$ & $156.82$ ($-4.5\%$) & $156.62$ ($-4.7\%$) \\
        $\Theta_4$ & $116.05$ & $72.05$ ($-37.9\%$) & $70.67$ ($-39.1\%$) \\
    \end{tabular}
    \caption{Comparison of resulting average per-period costs for Scenarios 1, 2, and 8 depending on the TPM of supply states. Relative savings compared to the deterministic approach (Scenario 1) in brackets.}
    \label{tab:sensitivity_supply}
\end{table}

The results presented in Table \ref{tab:sensitivity_supply} show a large variation in relative cost savings when comparing resulting average per period costs for under the deterministic approach to those obtained under full probabilistic information for different transition probability matrices on supply states. Due to the increased risk of (partial) supply shortages, in all cases considered here average total costs are higher than under the baseline matrix. This also leads to a decreased potential of reducing costs when incorporating probabilistic information for all sources of uncertainty (Scenario 8). However, while the low risk of supply shortages in the baseline case even increased total costs in Scenario 2, we find cost reductions for all cases in this analysis. Since lost sales are more expensive than inventory holding and shortage, a model incorporating knowledge of the TPM determines replenishment order quantities such that there is a larger safety stock. Therefore, comprehensive cost savings can be reached in Setting $\Theta_1$. A similar result can be obtained when considering $\Theta_4$ with a probability of 1/3 for all three supply states independent of the previous state. At the same time, savings in Settings $\Theta_2$ and $\Theta_3$ are much smaller. Due to the persistence of the same supply state, the retailer is rarely able to react to supply shortages by increasing the replenishment order quantity for the following period as there is still a large probability for shortages.

Finally, we consider a change in the cost structure for lost sales, inventory holding, and spoilage. As introduced above, in general, costs in e-grocery retailing are asymmetric due to economic consequences of lost sales. Hence the benefits of including additional information on probability distributions are expected to shrink if cost parameters become more symmetric. We test this hypothesis by changing the relationship between cost parameters. While assuming a constant relationship between inventory costs $v=0.1$ and spoilage costs $h=1$, we change the costs for one unit lost sales. In the first analysis, we assume that lost sales equal inventory costs leading to $b_1=0.1$. Furthermore, we consider $b_2=0.5$, $b_3=1$ (i.e.\ costs for lost sales and spoilage are identical), $b_4=2$ and $b_5=10$.

\begin{figure}[ht]
\centering
\includegraphics[width=0.7\textwidth]{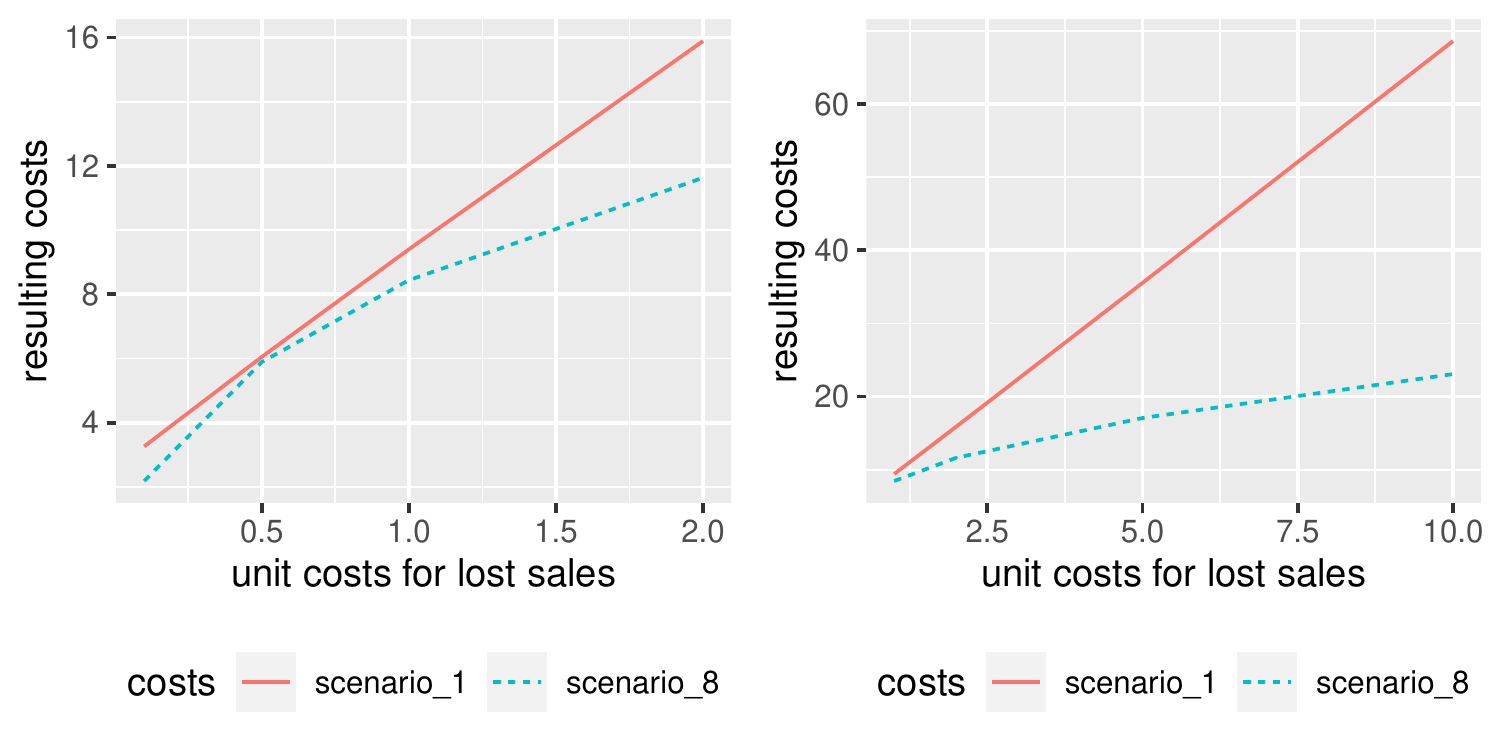}
\caption{Resulting average per period costs depending on unit costs for lost sales.}
\label{fig:sensitivity_cost_abs}
\end{figure}

Figure \ref{fig:sensitivity_cost_abs} shows average per period costs for the deterministic approach (Scenario 1, red line) and full probabilistic information (Scenario 8, blue line) for given unit costs for lost sales. If these costs are between unit costs for inventory holding and spoilage, the difference is negligible while it is more important to incorporate probability distributions if the cost structure is asymmetric.

\begin{figure}[ht]
\centering
\includegraphics[width=0.7\textwidth]{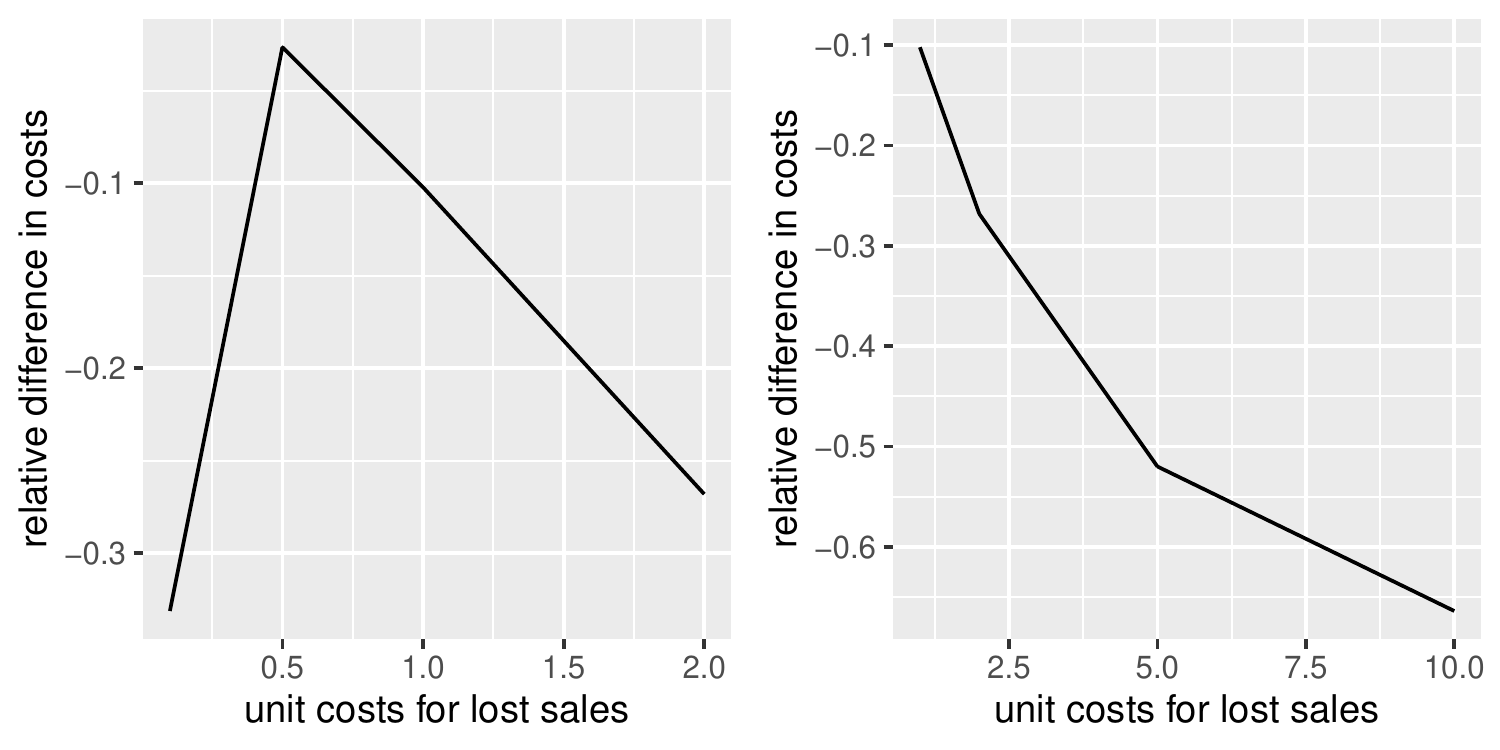}
\caption{Relative difference in resulting average per period costs between Scenarios 1 and 8 depending on unit costs for lost sales.} \label{fig:sensitivity_cost_rel}
\end{figure}

This result is confirmed by Figure \ref{fig:sensitivity_cost_rel} depicting the relative difference in average per-period costs. With $b=0.5$, i.e.\ costs per unit lost being half as high as costs per unit of spoilage, savings of only 2.6\% are achieved when including distributional information, whereas for the business case of e-grocery retailing with asymmetric cost structure (and corresponding high service-level targets) savings are much larger. As introduced above, for costs per unit lost of $b=5$, including information on the distribution of demand, spoilage, and supply shortages reduces costs by more than 50\%, while potential savings are even larger for $b=10$.

\end{spacing}
\end{document}